\documentclass[iop,apj,tighten,numberedappendix]{emulateapj}
\usepackage[breaklinks,colorlinks,urlcolor=blue,citecolor=blue,linkcolor=blue]{hyperref}

\usepackage{amsmath,amssymb}
\usepackage{cleveref}
\usepackage{graphicx}
\usepackage{bm}
\usepackage[toc,title,page]{appendix}
\usepackage{tocbibind}
\usepackage{xcolor,ulem}
\usepackage{color}

\usepackage{multirow}

\newcommand{\Msun}{{\rm M}_\odot}

\usepackage{lineno}

\def\lsim{~\rlap{$<$}{\lower 1.0ex\hbox{$\sim$}}}
\def\gsim{~\rlap{$>$}{\lower 1.0ex\hbox{$\sim$}}}

\shorttitle{Radio emission from QPEs}
\shortauthors{}

\begin{document}
\title{
Radio Emission Evolving with Quasi-Periodic Eruptions
}

\author{
Hiromichi Tagawa\altaffilmark{1} and 
Shigeo S Kimura\altaffilmark{2,3}
}

\affil{
\altaffilmark{1}Shanghai Astronomical Observatory, Shanghai, 200030, People$^{\prime}$s Republic of China\\
\altaffilmark{2}Astronomical Institute, Graduate School of Science, Tohoku University, Aoba, Sendai 980-8578, Japan\\
\altaffilmark{3}Frontier Research Institute for Interdisciplinary Sciences, Tohoku University, Sendai 980-8578, Japan
}
\email{E-mail: htagawa@shao.ac.cn}

\begin{abstract} 
A new class of transients, known as X-ray quasi-periodic eruptions (QPEs), has attracted considerable interest in the study of stellar and accretion disk evolution around supermassive black holes. 
Previous studies suggest that these eruptions can be effectively explained by flares resulting from collisions between stars and accretion disks. 
In one specific QPE system, J0230+28, a notable evolution of the radio flux has been reported at a frequency of $10~{\rm GHz}$, 
although an explanation for this phenomenon has not yet been provided. 
In this paper, we propose that the observed radio emission can be attributed to synchrotron radiation from shocks generated by collisions between gas ejected during interactions between a star and an accretion disk, and the winds emitted from the accretion disk. 
In this scenario, radio counterparts are easier to detect 
for events with longer recurrence times, 
as higher ejecta masses result in brighter radio emission. 
We estimate that the evolution of the radio flux spans several tens of days, 
constrained by synchrotron self-absorption. 
These predictions help delineate the timing of when the accretion disk gas becomes shocked and ejected, thereby contributing to the establishment of the QPE scenario. 
More densely sampled observations of the radio flux will be valuable for 
further constraining this scenario and 
for assessing intrinsic parameters such as plasma properties and the density profiles of winds and ejecta. 
\end{abstract}

\section{Introduction}

Recently, a new class of transients called X-ray quasi-periodic eruptions (QPEs) has been identified \citep[e.g.][]{Miniutti2019,Giustini2020,Arcodia2021}. 
QPEs are characterized by periodic flares with X-ray luminosities ranging from $L_{\rm X}\sim 10^{42}$ to $10^{43}~{\rm erg/s}$ and exhibit thermal-like spectra with temperatures around $\sim 100~{\rm eV}$. Each flare lasts $\sim 1$--$10$ hours, followed by a quiescent phase with 
a duty cycle of about $\sim 10\%$. 
While the periodicity of these eruptions is generally quite regular, 
some variations and evolutionary changes have been observed \citep{Miniutti2022,Arcodia2022,Arcodia2024_QPE2,Chakraborty2021,Chakraborty2025_stime,HernandezGarcia2025,Chakraborty2026,Arcodia2026,Guo2026}. 
To date, about ten QPE events have been reported, with emission originating from the central regions of (post-)active low-mass galaxies \citep{Wevers2022,Wevers2024,Gilbert2024,Shepherd2026ApJ}.  
Notably, many QPE events,
such as GSN069, XMMSL1~J024916.6-041244, eRO-QPE3, Tormund, AT2019qiz, AT2022upj, and Ansky, 
indicate relics or signatures that occurred following flares interpreted as tidal disruption events \citep{Miniutti2019,Chakraborty2021,Miniutti2022,Miniutti2023_reb,Quintin2023,Nicholl2024,Bykov2025,Chakraborty2025,Zhu2025,WuTao2026}.

\begin{figure*}
\begin{center}
\includegraphics[width=180mm]{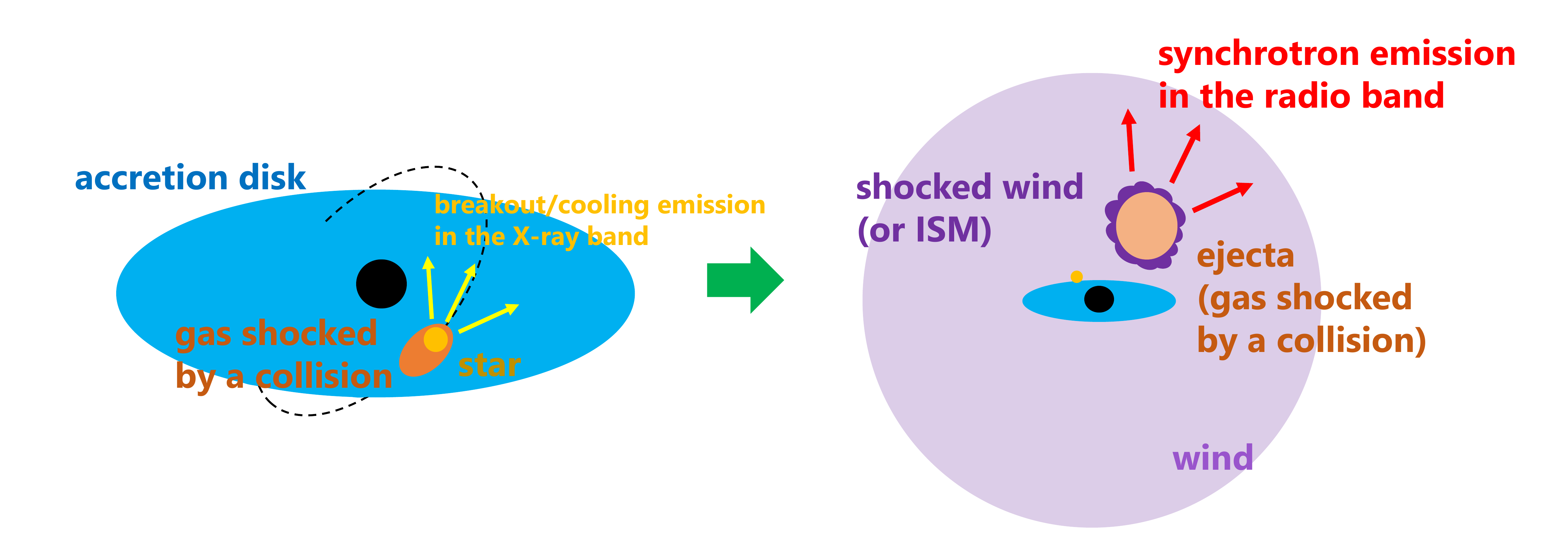}
\caption{
The schematic picture for the synchrotron emission produced in our model. 
X-ray QPEs arise from breakout emission resulting from shocks generated by collisions between a star and an accretion disk. 
Subsequently, radio flares are produced by synchrotron emission from shocks caused by collisions between gas ejected due to these collisions and the wind launched from the disk. 
}
\label{fig:schematic}
\end{center}
\end{figure*}

A promising model suggests that 
shocks from collisions between an accretion disk and a star could be the source of emission 
\citep{Nayakshin2004,Dai2010,Xian2021,Sukova2021,Lu2022,Franchini+2023,Ressler2024,Kaur2024_FP,Kejriwal2024,Miniutti2024,Zhou2024a,Zhou2024b,Chen2024_pop,WangY2024_TDE_AGN,Guolo2025_disk,Mummery2025,Suzuguchi2026,Mondek2026,Allievi2026,Linial2025,ChenBoAn2026,ZhangZirui2026,Chakraborty2026_warp}. 
One signature of this model is the regular patterns of strong-weak and long-short alternations in light curves \citep{Miniutti2022,Chakraborty2021}, likely originated from varying influences during two collisions per orbit for an eccentric star. 
These variations may arise from Doppler effects \citep{Tagawa2023,Miniutti2024}, differences in accretion disk density, the star's velocity \citep{Linial2023}, or asymmetric ejection of shocked gas \citep{Jankovic2026,Huang2025,Huang2026,LiuKun2026}. 
As emission processes, \citet{Linial2023} and \citet{Tagawa2023}, respectively, proposed that shock cooling and breakout emission could explain the properties of QPEs. 
Numerous simulations have investigated the properties of emission and shocks \citep{Vurm2024,Yao2024,Yao2026,Huang2025,Jankovic2026,Jankovivc2026b,LiuKun2026,Pelle2026,Huang2026,ZhuZhaohuan2026}. 
In shock cooling emission, the luminosity and the radiation temperature are expected to decrease continuously during this phase due to adiabatic expansion 
\citep{Vurm2024}. Increases and decreases in the radiation temperature before and after the peak luminosity have been confirmed by detecting the Wien tails \citep{Miniutti2019,Chakraborty2024,Guolo2024,Quintin2023,Nicholl2024}
and luminosity in the high-energy range of $\gtrsim$ a few $~{\rm keV}$ has been observed to be low in several events 
\citep{Giustini2020,Arcodia2021,Arcodia2024_QPE34}, posing a challenge to the shock cooling emission hypothesis. 
\citet{Linial2025} and \citet{Yao2026} suggested that a stream surrounding a star could explain the observed relation between the duration and recurrence time, 
while \citet{ZhuZhaohuan2026} found that realistic disk structures could produce this relation by breakout emission.

Associated with QPEs, radio emission has been reported to originate from galactic nuclei \citep{Shu2017,Miniutti2019,YangLei2022,Giustini2024,Guolo2024,Goodwin2025}. 
This radio emission could originate from jets produced from a central SMBH, outflows from an accretion disk, tidal disruption events, or gas ejected due to collisions between a star and an accretion disk. 
\citet{Linial2026} estimated the properties of radio emission caused by ejecta launched during the lifetime of QPE systems, especially focusing on radio emission associated with tidal disruption events, brightening long after the X-ray QPE phases. 

In this paper, we focus on 
the time evolution of radio emission during or shortly after the X-ray QPE phases. 
We particularly discuss scenarios for radio emission associated with X-ray flares in the QPE event, Swift~J0230+28 \citep{Evans2023,Guolo2024}. 
This event exhibits unique characteristics among QPEs, such as a slow rise in eruptions with a long duration of $\sim 4$ days, and a long recurrence time of 22 days. 
\citet{Guolo2024} monitored X-ray fluxes for about 11 cycles, with the 1st-6th and 10th-11th cycles being bright, while the 7th--9th cycles were dim. 
The radio emission was detected only once by the Very Large Array, 
with a flux of $F=93\pm 7~{\mu}{\rm Jy}$ at a frequency of $f=10^{10}~{\rm Hz}$ on ${\rm MJD}~59842$. 
The timing of this detection roughly coincides with the fifth cycle among the active phases observed by \citet{Guolo2024}. 
There are two additional constraints for non-detection, with flux values of $F\lesssim 15~{\mu}{\rm Jy}$ and $F\lesssim 25~{\mu}{\rm Jy}$ observed 80 days before and 91 days after the radio detection (${\rm MJD}~59762$ and ${\rm MJD}~59933$), respectively. 
The first non-detection corresponds to about ten days after the first cycle of X-ray QPEs, while the second non-detection occurs about seventy days after the sixth cycle of X-ray QPEs. 
Currently, the mechanisms potentially responsible for this radio emission have not been extensively explored.

We investigate whether this emission can be attributed to 
non-thermal radiation generated by shocks resulting from collisions between gas ejected by interactions between a star and an accretion disk, and the winds launched from either the accretion disk or from previous QPE episodes. 
The paper is organized as follows: 
\S~\ref{sec:method} details the methods used in our models,  
\S~\ref{sec:result} presents our main findings, 
\S~\ref{sec:discussions} discusses about X-ray observations of the event, Swift~J0230+28, 
and 
\S~\ref{sec:conclusions} summarizes our key conclusions.

\begin{table*}
\begin{center}
\caption{Fiducial values of our model parameters. 
}
\label{table:parameter_model}
\hspace{-5mm}
\begin{tabular}{c|c}
\hline 
Parameter & Fiducial value \\
\hline\hline
Conversion efficiency of shock kinetic energy to magnetic/electron energy & $\epsilon_{\rm B}=0.05$, $\epsilon_{\rm e}=0.2$\\\hline
Fraction that the shocked electrons is accelerated to be non-thermal electrons& $f_{\rm e}=0.1$\\\hline
Power-law slope for injected electrons accelerated by the first-order Fermi process& $p_{\rm e}=3$\\\hline
Ejecta velocity & $v_{\rm ej}=0.06~c$\\\hline
Ejecta mass & $m_{\rm ej}=4\times 10^{-4}~\Msun$\\\hline
Outflow rate of the wind & ${\dot M}_{\rm wind}=0.001~{\dot M}_{\rm Edd}$\\\hline
Power-law exponent for the ejecta density profile & $n=7$\\\hline
Power-law exponent for the wind density profile & $s=2$
\\\hline
\end{tabular}
\end{center}
\end{table*}

\section{Methods}

\label{sec:method}

We consider the following scenario for producing radio emission (see Fig.~\ref{fig:schematic} for a schematic illustration): 
In this event, X-ray QPEs are generated by emission from shocks resulting from collisions between a star and the accretion disk \citep{Linial2023,Franchini+2023,Tagawa2023}. 
Following these collisions, 
the shocked gas gains kinetic energy and 
is expelled from the disk. 
This expelled gas may be dominated by either shocked disk gas or stellar envelope material, depending on parameters, particularly the surface density of the accretion disk \citep{Yao2024}. 
This expelled gas may then collide with winds launched from the accretion disk, or with the interstellar medium, if the winds are weakly present. 
Additionally, it is possible for this ejected material to collide with ejecta from previous QPE episodes. 
These subsequent collisions can induce further shocks in the winds or interstellar medium, producing to non-thermal emission, similar to afterglow emission in gamma-ray bursts \citep[e.g.,][]{Meszaros1997,Sari1998}. 
This non-thermal emission, particularly synchrotron emission, can be bright in the radio bands and may explain the observed radio emission in Swift~J0230+28. 
In the following sections, we outline the methods used to calculate the properties of this non-thermal emission.

\subsection{Non-thermal emission}

We evaluate the properties of non-thermal emission resulting from forward shocks in 
winds launched from an accretion disk (or in the interstellar medium) 
that collide with ejected material. 
We utilize the formulas from \citet{Matsuoka2024}
to calculate the properties of synchrotron emission. 

In these shocks, electrons are accelerated via the first-order Fermi process, resulting in a power-law distribution of Lorentz factors in the fluid comoving frame $\gamma_{\rm e}$, such that $N(\gamma_{\rm e})d\gamma_{\rm e} \propto \gamma_{\rm e}^{-p_{\rm e}}d\gamma_{\rm e}$ with the power-law index $p_{\rm e}$. 
Plasma and/or magnetohydrodynamic instabilities also amplify the magnetic field. 
We assume that a fraction $f_{\rm e}$ of the shocked electrons is accelerated to become non-thermal electrons, 
while fractions $\epsilon_{\rm e}$ and $\epsilon_{\rm B}$ of the kinetic energy density dissipated at the shock are converted into non-thermal electrons and magnetic fields, respectively. 
Then, the minimum Lorentz factor for injected electrons is given by 
\begin{align}
\label{eq:gamma_min}
    \gamma_{\rm m}=\frac{p_{\rm e}-2}{p_{\rm e}-1}\frac{\epsilon_{\rm e}}{f_{\rm e}}\frac{m_{\rm p}v_{\rm sh}^2}{m_{\rm e}c^2}
\end{align}
where $m_{\rm p}$ and $m_{e}$ are the masses of the proton and electron, $v_{\rm sh}$ is the shock velocity, and $c$ is the speed of light. 
Eq.~\eqref{eq:gamma_min} ensures the conservation of energy and electron number. 
As cooling mechanisms for non-thermal electrons, 
we consider 
synchrotron emission, inverse Compton emission, 
adiabatic expansion, and Coulomb interactions. 
Synchrotron self-absorption is accounted for by assessing the emissivity and the absorption coefficient, and the optical depth to synchrotron self-absorption \citep{Rybicki1979}.

We predict the spectral energy distribution for synchrotron-self Compton and 2nd-order inverse Compton emission using Eqs. (33)--(36) in \citet{Fan2008}. 
We also estimate the external inverse Compton scattering of disk emission, assuming that a disk luminosity of $4\times 10^{41}~{\rm erg/s}$ at a frequency of $\nu_{\rm disk}\sim 3\times 10^{16}~{\rm erg/s}$, reflecting the SMBH mass and accretion rates of the disk. 
In the Klein-Nishina regime, $g\equiv \gamma_{\rm e}h\nu_{\rm disk}/m_{\rm e}c^2 \geq 1$, where $h$ is the Planck constant, 
we reduced the flux by a factor of $(1+g^2)$.

We incorporate free-free absorption as described by \citet{Matsuoka2020}, setting the thermal electron temperature of the ambient medium to $T_{\rm e}=10^5~{\rm K}$, as is commonly adopted for Type~II supernovae.

\subsection{Shock components and evolution}

In this section, we describe the distributions of ejecta and winds, as well as the evolution of shocks.

We assume that the density distribution of the wind is given by 
\begin{align}
\rho_{\rm wind}(r)=\frac{{\dot M}_{\rm wind}}{4 \pi v_{\rm wind} {\rm max}(r_*,r)^2} 
\left(\frac{r_*}{{\rm max}(r_*,r)}\right)^{s-2}, 
\end{align}
where ${\dot M}_{\rm wind}$ is the outflow rate of the wind, 
$v_{\rm wind}$ is the wind velocity, 
$r_*$ is the distance from the central SMBH to the location where the collision between the star and the disk occurs, 
and $r$ is the distance from the collision location to the ejecta. 
The case with $s=2$ represents successful winds, 
while $s>2$ indicates cases where the wind density steeply declines beyond $r=r_*$, possibly consisting of failed winds or previous ejecta.
We also consider scenarios where the winds are not efficiently launched, leading to the ejecta to colliding with the interstellar medium, which is characterized by a constant density ($\rho_{\rm ISM}={\rm const}$).

Additionally, 
we assume that the ejecta density profile follows a power-law form, $\rho_{\rm ej}\propto r^{-n}$, even though it may not be a precise approximation (see also \S\ref{sec:parameters}).

While the swept mass ($m_{\rm coll}(r)=4\pi \int^r_0 \rho_{\rm wind}(r')r'^2 dr'$) is less than the ejecta mass ($m_{\rm ej}>m_{\rm coll}$), 
we assume that the shock adiabatically decelerates according to the relation $R_{\rm sh} \propto t^{n-3/n-s'}$ \citep{Chevalier1982} following the formulae in \citet{Matsuoka2024}, 
where $s'$ is the power-law index for the wind density profile at $r$, as $\rho_{\rm wind}\propto r^{-s'}$. 
We assume that the post-shocked density of the wind is four times larger than that of the preshocked wind \citep{Matsuoka2020}. 
Once the swept mass exceeds the ejecta mass, 
the shock transitions into the Sedov-Taylor phase, characterized by $R_{\rm sh} \propto t^{2/(5-s')}$. 
During this transition of the deceleration phases or the wind profile, 
we assume that the radius and the velocity evolve without any discontinuous changes.

\subsection{Parameters}

\label{sec:parameters}

This section outlines the fiducial values for the model parameters, as also listed in Table~\ref{table:parameter_model}.

We adopt 
the fractions of post-shock energy carried by the post-shock magnetic field 
and by electrons as $\epsilon_{\rm{B}}=0.05$ 
and $\epsilon_{\rm{e}}=0.2$ \citep{Panaitescu2001,Frail2005}, respectively. 
These parameters are adjusted 
to roughly reproduce the radio flux peak suggested for Swift J0230+28 (Fig.~\ref{fig:l_ev_para}). 
Additionally, we set the power-law slope for injected electrons accelerated by the first-order Fermi process to $p_{\rm e}=3$, motivated by values estimated for non-thermal emission from non-relativistic shocks (e.g. \citealt{Chevalier2006,Soderberg2012}). 
However, caution is required, especially at high-energy tails, as such a steep profile may arise from other components (such as a supra-thermal component) and a shallower slope ($p_{\rm e}\sim 2$) may typically be realized for electron Lorentz factors of $\gtrsim 100$ \citep{Maeda2013,Caprioli2014}.

We assume that the ejecta density profile follows $n=7$. 
This value is commonly used for supernovae ejecta \citep{Chevalier1982} and is roughly consistent with the QPE ejecta profile predicted by simulations over a limited range, although it also resembles an exponential profile (e.g., Fig.~13 in \citealt{Huang2026}). The exact values would vary depending on the origin of the ejecta (disk gas or stellar envelopes, \citealt{Linial2025}), along with the specifics of the calculations, which will need to be clarified in future studies. 
For the wind density profile, we set $s=2$, which is a standard value for successful winds.

Assuming that flares occur twice per orbit around the SMBH,  
the semi-major axis of the stellar orbit is estimated by 
$r_*=(P_{\rm QPE}/\pi)^{2/3}(GM_{\rm SMBH})^{1/3}$, where $P_{\rm QPE}$ is the recurrence time between consecutive QPEs, $G$ is the gravitational constant, and 
$M_{\rm SMBH}$ is the SMBH mass. 
We adopt $r_* = 6\times 10^{14}~{\rm cm}$ reflecting 
$P_{\rm QPE}=22~{\rm d}$ and $M_{\rm SMBH}\sim 10^{6.6}~\Msun$, as estimated by \citet{Guolo2024}.  
At the Keplerian velocity at this location ($v_{\rm Kep}\sim 0.03~c$), 
and assuming the counter-rotation predicted by our previous study, 
we set the ejection velocity of the ejecta to $v_{\rm ej}\sim 2~v_{\rm Kep}\sim 0.06~c$. 
We assume that the velocity of the winds launched from the surface of the accretion disk is $v_{\rm wind}\sim \eta_{\rm vw} v_{\rm Kep}\sim 0.01~c (\eta_{\rm vw}/0.3)$, where $\eta_{\rm vw}$ accounts for variations in wind velocity. 
We set $\eta_{\rm vw}\sim 0.3$, to ensure the wind velocity is slower than the ejecta velocity, considering failed winds \citep{Wada2012,Williamson2020,Kudoh2023}, which allows for strong shocks. This condition is necessary for applying the shock evolution formulae \citep{Chevalier1982}. 
This is an optimistic choice for the shock kinetic energy, compared to successful winds of $\eta_{\rm vw}\sim 1$.

We set the inflow rate of the disk to be ${\dot M}_{\rm disk}=0.001~{\dot M}_{\rm Edd}$, which is derived from the AGN luminosity of $\sim 4\times 10^{41}~{\rm erg/s}$ \citep{Pasham2024_Swift}, 
considering a bolometric correction factor of 10 \citep{Duras2020} and the SMBH mass, 
where ${\dot M}_{\rm Edd}=L_{\rm Edd}/\eta_{\rm rad}c^2$ is the Eddington accretion rate,  $L_{\rm Edd}$ is the Eddington luminosity for the SMBH, 
and $\eta_{\rm rad}$ is the radiative efficiency, which is adopted to be $\eta_{\rm rad}\sim 3\%$, reflecting the adopted accretion rate \citep{Xie2012}. 
Note that for the fiducial parameters, the inner truncation radius for a thin disk ($\sim 4\times 10^{14}~{\rm cm}$) is barely smaller than the location of collisions, according to the ADAF model (the most preferred model) in \citet{Czerny2004}. 
We set ${\dot M}_{\rm wind}= \eta_{\rm mw}{\dot M}_{\rm disk}$, 
where $\eta_{\rm mw}$ represents the fraction of wind mass loss compared to accretion. We set $\eta_{\rm mw}=0.3$ 
assuming efficient wind mass loss from the disk. 
Note that since the properties of the radio emission are influenced by $\rho_{\rm wind}(r)$, the effects of $v_{\rm wind}$ and ${\dot M}_{\rm wind}$ are degenerate.

Utilizing the properties mentioned above, 
we derive the disk surface density to be $\Sigma_{\rm disk}\sim 3\times 10^{4}~{\rm g/cm^2}$, assuming a viscous parameter of $\alpha=0.02$ 
and considering the free-free absorption opacity expected at $r_*\sim 10^3~r_{\rm g}$ \citep{Haiman2009}. 
We assume a stellar envelope expands to the Hill radius $R_{\rm Hill}=2.6\times 10^{12}~{\rm cm}$, 
motivated by the heating of the stellar envelope due to collisions \citep{Yao2024}. 
We simply assume that the ejecta mass is given by 
the mass swept up by a vertical collision as $m_{\rm ej} \sim  \pi R_{\rm Hill}^2 \Sigma_{\rm disk}\simeq 4\times 10^{-4}~{\Msun}$ (see \citealt{Chakraborty2026_outflow} for implications related to another event).

\begin{figure}
\begin{center}
\includegraphics[width=90mm]{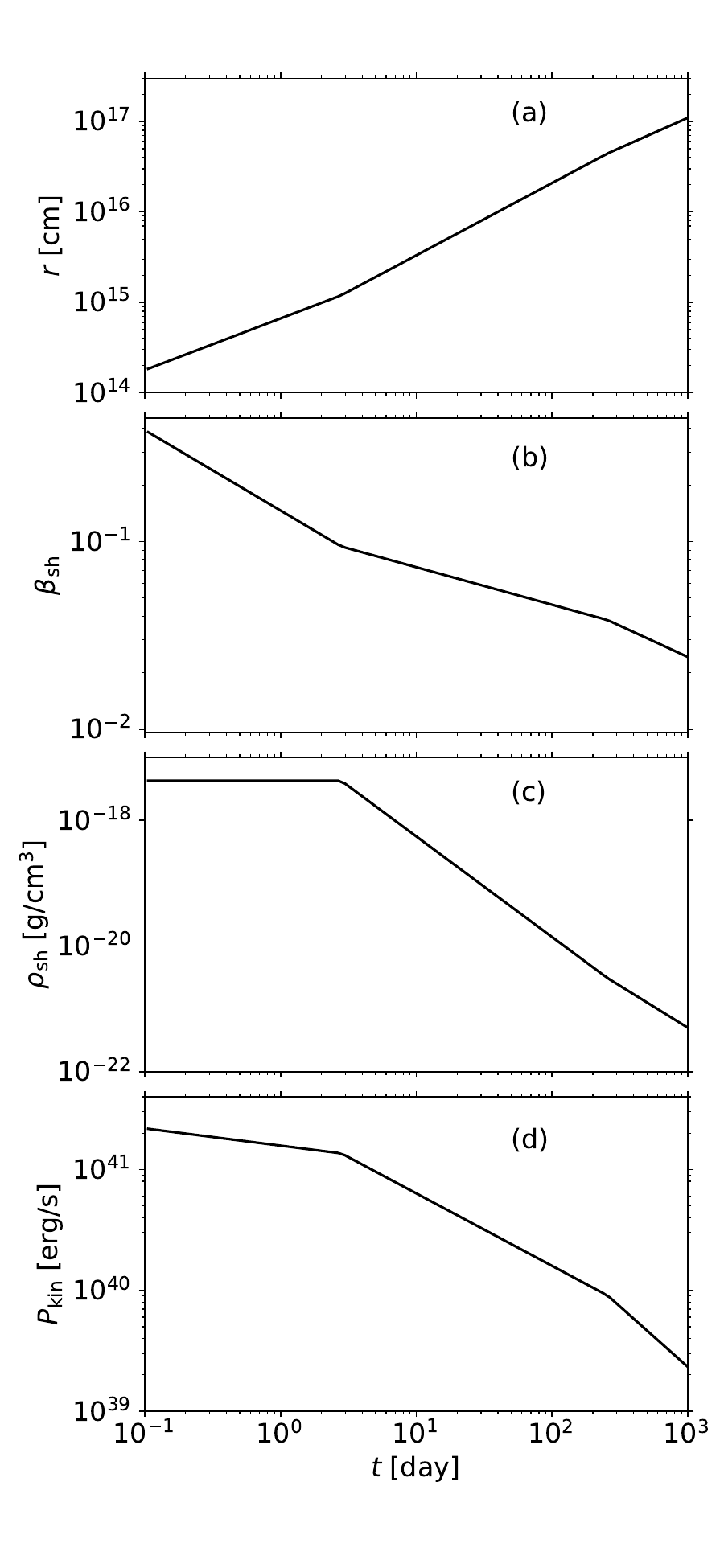}
\caption{
The evolution of shocks for the fiducial model. 
The panels~(a), (b), (c), and (d) show the size, velocity, and density, and power of shocked gas. 
}
\label{fig:l_ev_fid}
\end{center}
\end{figure}

\begin{figure*}
\begin{center}
\includegraphics[width=185mm]{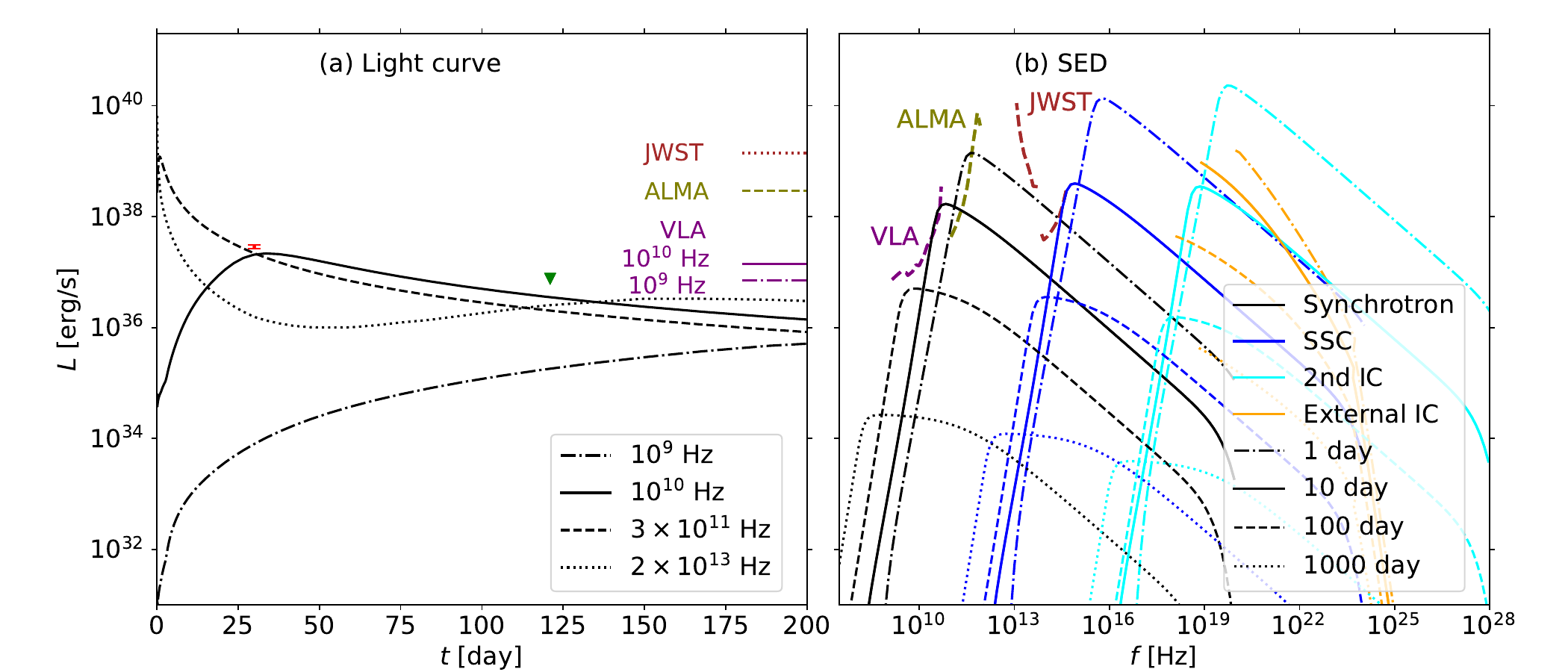}
\caption{
The panel~(a) shows shows the light curves for the fiducial model at four frequency bands: $f=10^9~{\rm Hz}$ (dash-dotted), $10^{10}~{\rm Hz}$ (solid), $3\times 10^{11}~{\rm Hz}$ (dashed), and $2\times 10^{13}~{\rm Hz}$ (dotted). The red and green points represent the radio detection and the upper limit 
at ${\rm MJD}~59842$ and ${\rm MJD}~59933$, 
respectively. 
The timing of the radio detection is noted to occur at $t=30~{\rm days}$. 
The panel~(b) shows the spectral energy distribution for the fiducial model at four phases, $t=1$ (dash-dotted), $10$ (solid), $10^2$ (dashed), and $10^3~{\rm days}$ (dotted). 
The black, blue, cyan, and orange lines represent 
the distributions for synchrotron, synchrotron-self Compton, 
2nd-order inverse Compton, and external inverse Compton emission, respectively. 
The dashed-dotted and solid purple lines represent 
the sensitivities of the Very Large Array (VLA) at $10^9$ and $10^{10}~{\rm Hz}$ with an integration time of $t_{\rm int}=1~{\rm h}$ \citep{DiFrancesco2019}, respectively. 
The dashed yellow line represents the Atacama Large Millimeter Array (ALMA) with $t_{\rm int}=10^4~{\rm s}$ \footnote{https://almascience.nao.ac.jp/proposing/sensitivity-calculator}. 
The dotted brown line represents the James Webb Space Telescope (JWST) with $t_{\rm int}=10^4~{\rm s}$ \citep{Rigby2023}, respectively. 
}
\label{fig:l_sed_fid}
\end{center}
\end{figure*}

\begin{figure*}
\begin{center}
\includegraphics[width=185mm]{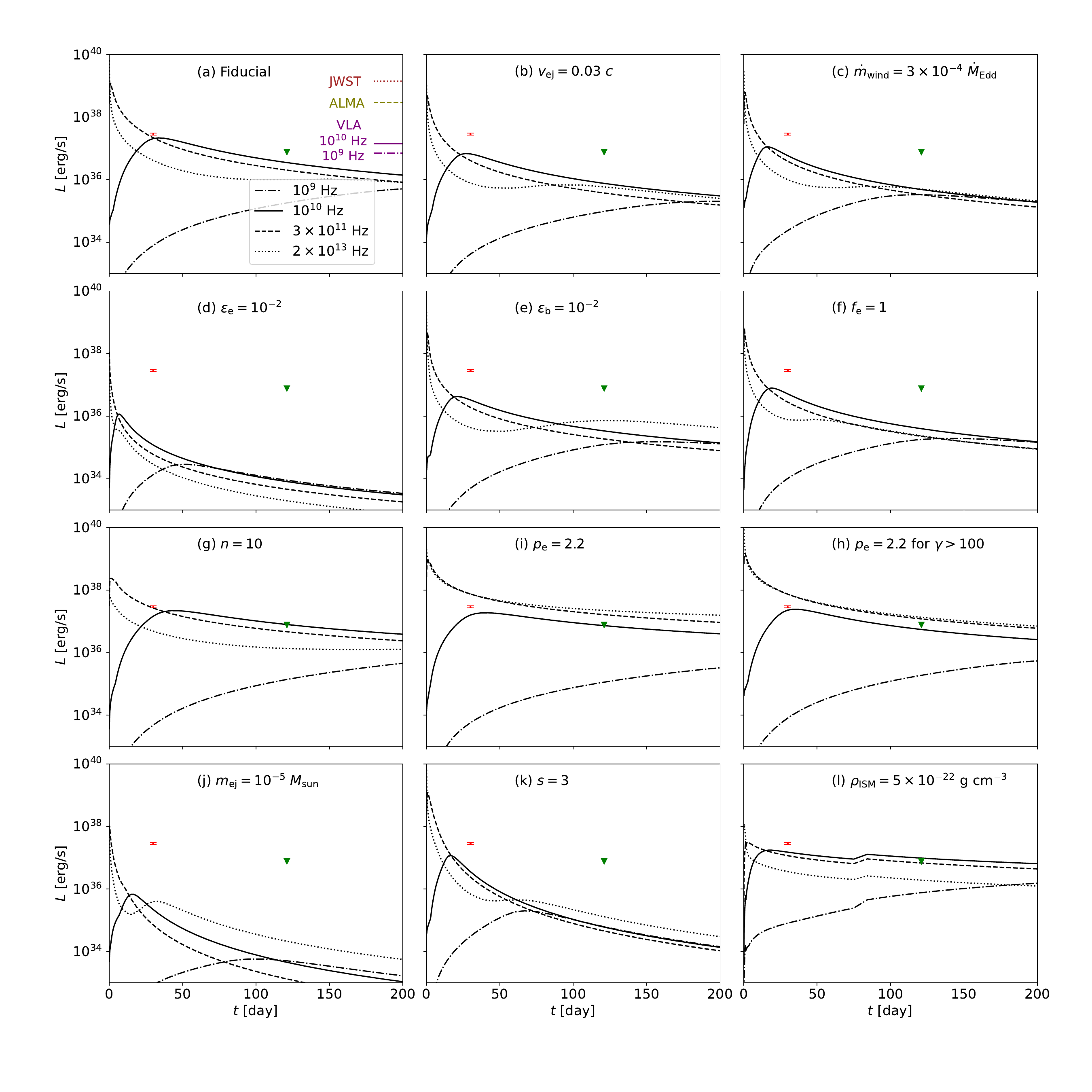}
\caption{
Same as the panel~(a) of Fig.~\ref{fig:l_sed_fid}, 
but the results for the fiducial model (panel~a), 
the model with $v_{\rm ej}=0.03~c$ (panel~b), 
${\dot m}_{\rm wind}=3\times 10^{-4}~{\dot M}_{\rm Edd}$ (panel~c), 
$\epsilon_{\rm e}=10^{-2}$ (panel~d), 
$\epsilon_{\rm B}=10^{-2}$ (panel~e), 
$f_{\rm e}=1$ (panel~f), 
$n=10$ (panel~g), 
$p_{\rm e}=2.2$ (panel~h), 
$p_{\rm e}=2.2$ for electron Lorentz factors of $\gamma_{\rm e}\geq 100$ (panel~i), 
$m_{\rm ej}=10^{-5}~\Msun$ (panel~j), 
$s=3$ (panel~k), 
and $\rho_{\rm ISM}=5\times 10^{-22}~{\rm g~cm^{-3}}$ (panel~l). 
To explain the observed radio fluxes, 
it is necessary for the shocked energy to be efficiently converted into radiation energy. 
}
\label{fig:l_ev_para}
\end{center}
\end{figure*}

\begin{figure*}
\begin{center}
\includegraphics[width=180mm]{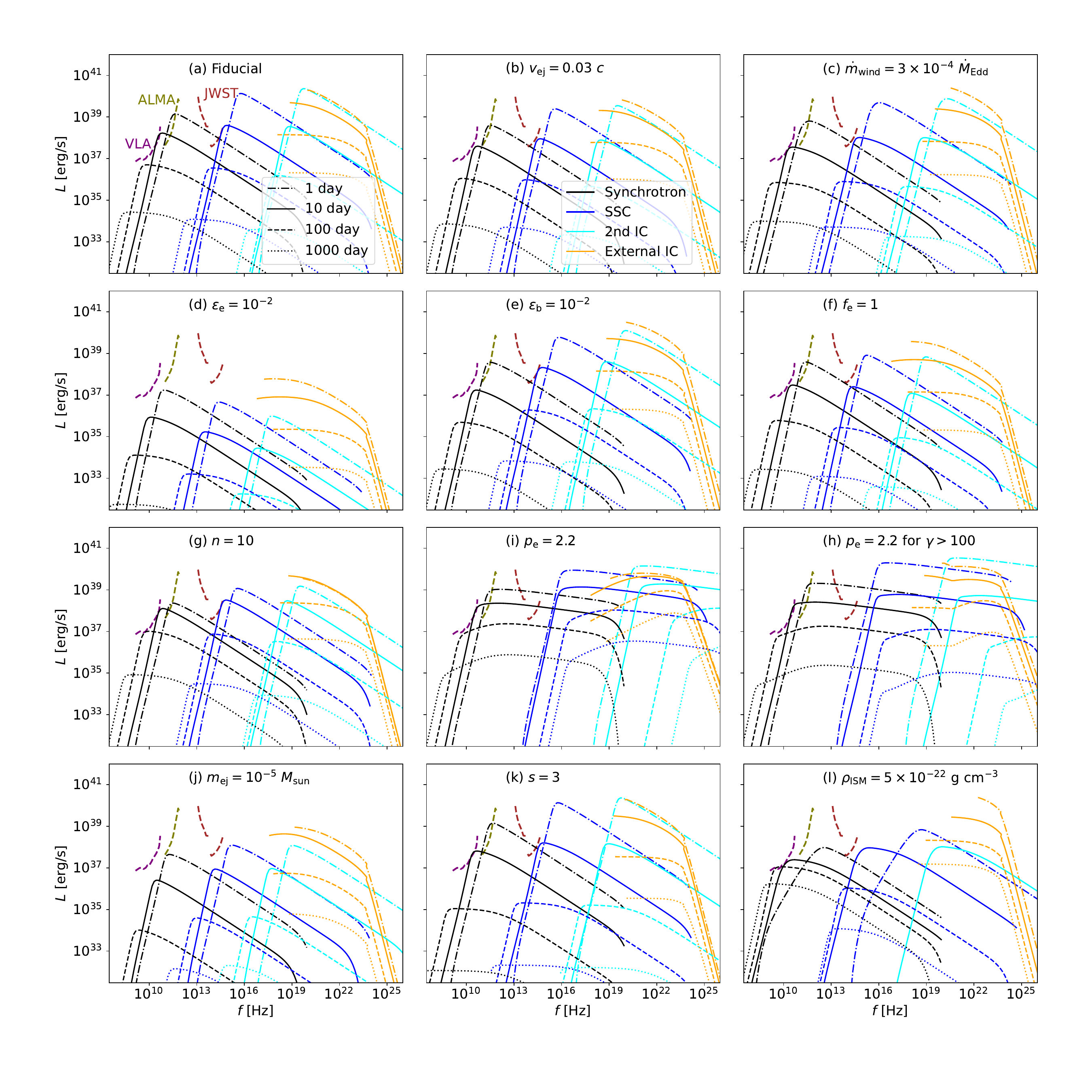}
\caption{
Same as the panel~(a) of Fig.~\ref{fig:l_sed_fid}, 
but results for the models shown in Fig.~\ref{fig:l_ev_para}. 
The spectrum is nearly flat above the absorption frequency and below the cooling frequency. 
}
\label{fig:l_sed_para}
\end{center}
\end{figure*}

\begin{figure}
\includegraphics[width=90mm]{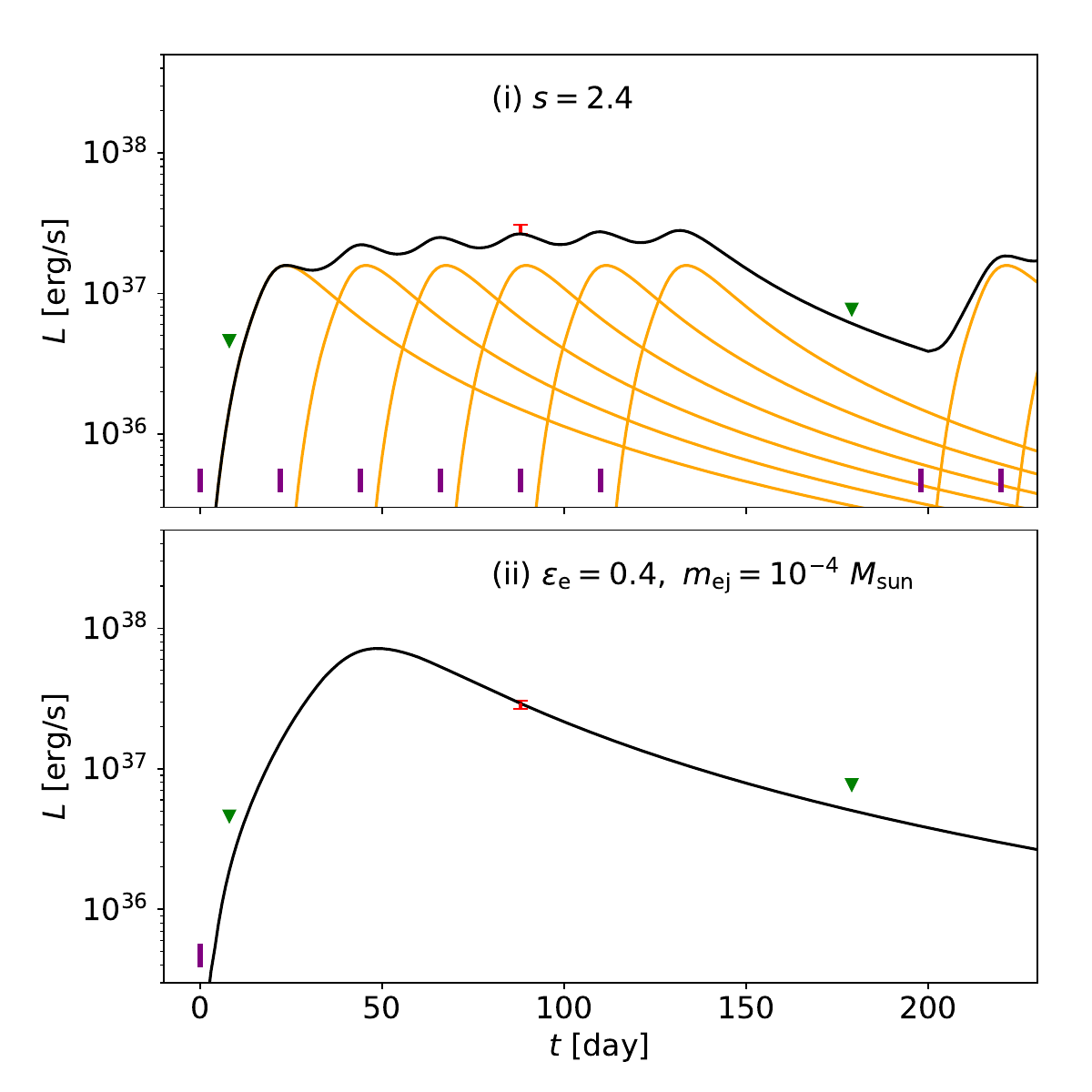}
\caption{
The light curves at 10 GHz depict two scenarios that explain the observed radio fluxes: 
(i) The radio emission is associated with the bright QPE flares in the 1st--6th, 10th, and 11th cycles. 
In this case, a steeper wind profile with $s=2.4$ is used to ensure a rapid decay after the 6th flare. The luminosity contributed by each flare and all flares combined is shown by the orange and black lines, respectively. 
(ii) The radio emission is associated only with the first QPE flare. In this scenario, the surrounding intermediate winds are evacuated by shocks resulting from collisions with ejecta produced during the first flare. 
A higher efficiency of $\epsilon_{\rm e}=0.4$ and $m_{\rm ej}=10^{-4}~\Msun$ is adopted to match the observed radio flux. 
A red error bar indicates the detected flux, while green triangles denote the upper limit of the flux. 
Purple lines mark the timing of QPE flares whose associated ejecta collide with surrounding winds, producing radio emission. 
These scenarios represent totally different light curves, which can be distinguished through more frequent and/or multi-wavelength radio observations. 
}
\label{fig:l_ev_obs}
\end{figure}

\section{Results}

\label{sec:result}

\subsection{Shock evolution and non-thermal emission}

We first describe how shocks evolve and produce non-thermal emission in the fiducial model. 
Fig.~\ref{fig:l_ev_fid} illustrates the evolution of the size, velocity, density, and power of the shocked gas. 
Initially, in the fiducial model, the shock propagates through a medium of constant density with $s'=0$ while experiencing significant deceleration (panels~b and c). 
During the early phase, 
the shock power slightly decreases over time (panel~d). 
The shock reaches the radius $r_*$ at $\sim 3~{\rm days}$, after which the density of the shocked wind begins to decline, 
transitioning to $s'=2$, 
resulting in a more pronounced decrease in shock power. 
The ejecta decelerates more effectively after $\sim 260~{\rm days}$ post-ejection, as it sweeps up a wind mass comparable to that of the ejecta mass.

Fig.~\ref{fig:l_sed_fid}~(a) shows the light curves across four frequency bands, $f=10^9~{\rm Hz}$ (dash-dotted), $10^{10}~{\rm Hz}$ (solid), $3\times 10^{11}~{\rm Hz}$ (dashed), and $2\times 10^{13}~{\rm Hz}$ (dotted), while Fig.~\ref{fig:l_sed_fid}~(b) presents the spectral energy distribution at four phases, $t=1~{\rm d}$ (dash-dotted), $10~{\rm d}$ (solid), $10^2~{\rm d}$ (dashed), and $10^3~{\rm d}$ (dotted). 
In the fiducial model, 
the frequency corresponding to the minimum Lorentz factor $(f_{\rm m}\equiv f_{\rm syn} \gamma_{\rm m}^2)$ exceeds the cooling frequency ($f_{\rm c}\equiv f_{\rm syn} \gamma_{\rm c}^2$) at $t\lesssim 1~{\rm d}$, 
leading to synchrotron emission in a fast cooling regime \citep[e.g.,][]{Fan2008}, 
where $f_{\rm syn}$ is the synchrotron frequency, 
and $\gamma_{\rm c}$ is the cooling Lorentz factor of electrons, above which electrons cool via radiation within dynamical timescale. 
For example, when synchrotron emission dominates cooling, 
$\gamma_{\rm c}=6\pi c m_{\rm e}/\sigma_{\rm T}B^2 t$, where $\sigma_{\rm T}$ is the Thomson scattering cross section, and $B$ is the magnetic field in the shocked ejecta. 
In the fast cooling regime, all the injected electrons are affected by cooling, resulting in a softer radiation spectrum. 
As time progresses, the shock decelerates and the minimum Lorentz factor decreases, transitioning to a slow cooling regime around $t\sim 1~{\rm d}$, 
in which the electrons with $\gamma_{\rm m}$ do not cool.

The peak energy, below which the luminosity steeply declines (Fig.~\ref{fig:l_sed_fid}~b), corresponds to the absorption frequency, $f_{\rm a}\equiv f_{\rm syn} \gamma_{\rm a}^2$, where $\gamma_{\rm a}$ is the absorption Lorentz factor, below which synchrotron self-absorption is effective. 
A caveat to our model is that 
it does not account for heating due to synchrotron self-absorption. 
This effect can increase the peak luminosity at the absorption frequency if that frequency is higher than the minimum and cooling frequencies \citep{Gao2013,Tagawa2023_breakout}.

During the deceleration of shocks, 
the shock luminosity decreases. 
In the early phase ($t\lesssim 22~{\rm days}$), when synchrotron emission dominates radiative cooling, this reduction in shock luminosity decreases the magnetic field, which lowers the cooling frequency. 
In the later phases, where external inverse Compton scattering becomes the dominant form of radiative cooling, the increase in the shocked radius reduces the photon density, lowering the cooling frequency. 
Additionally, the absorption frequency decreases because 
the reduction in shock luminosity lowers both the non-thermal electron density and the magnetic field, reducing the optical depth for synchrotron emission. 
For example, at $t=1~{\rm days}$, 
the cooling frequency is $f_{\rm c}\simeq 8\times 10^{10}~{\rm Hz}$, 
the absorption frequency is $f_{\rm a}\sim 10^{12}~{\rm Hz}$ (Fig.~\ref{fig:l_sed_para}), and 
the frequency corresponding to the minimum Lorentz factor is $f_{\rm m}\simeq 1\times 10^{11}~{\rm Hz}$. 
Subsequently, at $t=100~{\rm days}$, the cooling frequency rises to $f_{\rm c}\sim 2\times 10^{11}~{\rm Hz}$, 
whereas the minimum and absorption frequencies decrease to 
$f_{\rm m}\sim 2\times 10^{7}~{\rm Hz}$ and 
$f_{\rm a}\lesssim 10^{10}~{\rm Hz}$. 
The evolution of $f_{\rm c}$ and $f_{\rm a}$ is depicted by the widening, flatter spectral regions in later phases (dashed lines in panel~b of Fig.~\ref{fig:l_sed_fid}).

The free-free absorption is only effective at $\lesssim 1~{\rm d}$ at $f\lesssim 10^{9}$--$10^{10}~{\rm Hz}$, roughly consistent with analytical estimates \citep[e.g.,][]{Fransson1998}. This absorption has negligible influence on light curves and the spectral energy distribution.

For frequencies $f\geq10^{12}~{\rm Hz}$, synchrotron self-absorption is ineffective throughout the evolution. In contrast, 
the luminosity at $f=10^{10}~{\rm Hz}$ peaks when the emission becomes optically thin to synchrotron self-absorption (panel~a of Fig.~\ref{fig:l_sed_fid}). 
The significant delay of $\sim 30~{\rm d}$ between the ejection of shocked gas and the peak luminosity in the observed band is caused by the evolution of the absorption frequency. This suggests the following: the radio emission observed in Swift~J0230+28 cannot originate from non-thermal emission produced by material ejected during the active phase of the X-ray QPE coinciding with the timing of the radio emission detection. 
Instead, the radio emission must come from material that was ejected at least a few QPE cycles earlier. 
Consequently, this constrains the scenarios that could produce the emission. 
These constraints apply whenever the observed radio flux originates from synchrotron emission produced by shocks generated from gas ejected by collisions that create QPEs.

The peak luminosities of the synchrotron-self Compton, 2nd-order Compton, and external inverse-Compton emission (blue, cyan, and orange lines in the panel~b of Fig.~\ref{fig:l_sed_fid}) are typically comparable compared to or somewhat brighter than that of synchrotron emission. 
The ratios of these peak luminosities are primarily influenced by the ratio between the plasma parameters and the magnetic and photon density. 
Their peak frequencies fall within the infrared to gamma-ray bands.

2nd-order Compton and external inverse-Compton emission is bright in X-ray to gamma-ray bands, which is challenging to detect due to their lower flux ($\lesssim 3\times 10^{-16}~{\rm erg~s^{-1}~cm^{-2}}(L/10^{39}~{\rm erg/s})$, assuming a luminosity distance of $165~{\rm Mpc}$), considering lower sensitivities of telescopes in these bands.

Synchrotron and synchrotron-self Compton emission 
in the infrared band may be detectable by the James Webb Space Telescope (JWST), especially in the early phases, depending on the parameters (brown dashed lines in Fig.~\ref{fig:l_sed_fid}~b). 
Detection by JWST might be hindered by contamination from dust emission associated with active galactic nuclei \citep[AGN,][]{Lyu2022}, if QPEs are linked to AGNs with dusty tori. However, QPEs are generally associated with tidal disruption events rather than AGNs, which may reduce this source of contamination. 
Contamination from the host galaxy is expected to be subtracted, since its infrared variability is modest. 
At $f=2\times 10^{13}~{\rm Hz}$, the luminosity slightly increases around $\gtrsim 100~{\rm days}$. This due to synchrotron-self Compton emission, which is typically dimmer compared to early synchrotron emission (see also Fig.~\ref{fig:l_ev_para}).

Synchrotron emission might be observable by radio telescopes including the Very Large Array (VLA) and the Atacama Large Millimeter Array (ALMA). 
ALMA may detect radio emission across a wider range of parameter space (see purple and yellow lines in Fig.~\ref{fig:l_sed_para}), while the detection is only possible in the early phases (yellow line in Fig.~\ref{fig:l_sed_fid}~b). 
Detection in later phases is more suited to using the VLA at $10^{10}~{\rm Hz}$ (purple lines in Fig.~\ref{fig:l_ev_para}), as observed by \citet{Guolo2024}. 
Thus, non-thermal emission is observable from radio to infrared wavelengths. 
These multiwavelength observations are invaluable for constraining the physical parameters and identifying the scenarios that produce non-thermal emission and QPEs.

\subsection{Parameter dependence}

Fig.~\ref{fig:l_ev_para} and Fig.~\ref{fig:l_sed_para} 
show the light curves and the spectral energy distributions, respectively, similar to Fig.~\ref{fig:l_sed_fid}, but for 12 different models. 
In the different panels, we show various models: 
the fiducial model (panel~a), 
the model with $v_{\rm ej}=0.03~c$ (panel~b), 
${\dot m}_{\rm wind}=3\times 10^{-4}~{\dot M}_{\rm Edd}$ (panel~c), 
$\epsilon_{\rm e}=10^{-2}$ (panel~d), 
$\epsilon_{\rm B}=10^{-2}$ (panel~e), 
$f_{\rm e}=1$ (panel~f), 
$n=10$ (panel~g), 
$p_{\rm e}=2.2$ (panel~h), 
$p_{\rm e}=2.2$ for electron Lorentz factors of $\gamma_{\rm e}\geq 100$ (panel~i), 
$m_{\rm ej}=10^{-5}~\Msun$ (panel~j), 
$s=3$ (panel~k), 
and $\rho_{\rm ISM}=5\times 10^{-22}~{\rm g~cm^{-3}}$ with no winds (panel~l).

We observe that the peak flux at $10~{\rm GHz}$ becomes dimmer for low $v_{\rm ej}$ (panel~b), 
low ${\dot m}_{\rm wind}$ (panel~c), 
low $\epsilon_{\rm e}$ (panel~d), 
low $\epsilon_{\rm B}$ (panel~e), 
high $f_{\rm e}$ (panel~f), 
high $p_{\rm e}$ (panel~h), 
low $m_{\rm ej}$ (panel~j), 
and high $s$ (panel~k). 
The parameters $v_{\rm ej}$, 
${\dot m}_{\rm disk}$, 
and $m_{\rm ej}$ are related to the shock kinetic energy density. 
The plasma parameters $\epsilon_{\rm e}$, $\epsilon_{\rm B}$, $f_{\rm e}$, 
and $p_{\rm e}$ influence the conversion efficiency of the kinetic energy into radio flux \citep[e.g.][]{Fan2008}. 
The parameters $\epsilon_{\rm e}$, $\epsilon_{\rm B}$, and $f_{\rm e}$ also affect the minimum Lorentz factor of the non-thermal electrons 
and the synchrotron luminosity at the peak frequency.

In the model presented in panel~(l) of Fig.~\ref{fig:l_ev_para}, 
the ambient medium density is set to a constant value of $\rho_{\rm ISM}=5\times 10^{-22}~{\rm g~cm^{-3}}$. 
In this setup, 
it is assumed that winds are not efficiently produced by the accretion disk 
and that an interstellar medium is present \citep[e.g.,][]{Launhardt2002}. 
The deceleration of shocks is more efficient, allowing for an earlier peak in the emission at $10^{10}~{\rm GHz}$. 
The sudden change in luminosity at $\lesssim 100~{\rm d}$ is due to the disappearance of 2nd-order inverse Compton emission, which appears if ${\rm max}\left(\gamma_{\rm m}^3 f_{\rm m},\gamma_{\rm c}^3 f_{\rm c}\right)<m_{\rm e} c^2/2$ \citep{Fan2008}.

The parameters $p_{\rm e}$, $m_{\rm ej}$, $s$ and $n$ influence the decay timescale of the synchrotron emission. 
We observe that a lower value of $p_{\rm e}$ (panel~h of Fig.~\ref{fig:l_ev_para}) leads to a shallower decay \citep[e.g.,][]{Sari1998}. 
With a lower value of $m_{\rm ej}$ (panel~j), 
the decay of the flux becomes faster. 
This occurs because for $m_{\rm ej}= 10^{-5}~\Msun$, 
the transition to the Sedov-Taylor phase occurs at $11~{\rm d}$, 
efficiently decreasing the shock power. 
For larger values of $s$, 
the wind density declines sharply as shocks proceed, 
resulting in a more rapid decrease in the wind density. 
Consequently, the decay of the shock kinetic energy and the flux occurs more quickly. 
For larger values of $n$, 
shocks evolve with $R_{\rm sh}\propto t^{n-3/n-s}$, 
leading to a slower decline in both the kinetic power and the flux.

If the decay of the radio flux is slow, the models cannot adequately explain the detection of the radio flare and its subsequent non-detection after 91 days from the initial observation. 
In the fiducial model, 
the radio flux decreases by a factor of $\sim 4$ 
over a period of $\sim 60~{\rm days}$. This behavior may be consistent with observations if the radio flux peaks around the time of the detection. 
More detailed scenarios are discussed in the next subsection.

\subsection{Possible scenario}

\label{sec:scenario}

From the relationship between the bright X-ray flares and the radio observations, 
we can constrain the scenarios that produce radio emission. 
If synchrotron emission is regularly generated in association with the ejection of gas during stellar collisions with the accretion disk in every cycle (every $\sim 22$ days), 
the non-detections on ${\rm MJD}~59762$ and ${\rm MJD}~59933$ cannot be accounted for by the slow evolution of the emission, as 
illustrated in Fig.~\ref{fig:l_ev_para}. 

Notably, \citet{Pasham2024_Swift} reported that X-ray QPEs are absent 
$\sim 100$--$200~{\rm days}$ before and after the active X-ray phases described by \citet{Guolo2024}. 
This suggests that the processes generating X-ray QPEs and radio emission may be 
intermittent 
rather than static. 
Even during the active phases observed by \citet{Guolo2024}, 
there are low X-ray luminosity cycles during the 7th, 8th, and 9th cycles. 

Assuming that bright synchrotron emission is produced alongside mass ejection during the cycles with bright QPEs (the 1st-6th and 10th-11th cycles) but is not generated during the phases prior to the first X-ray QPE cycle observed by the Neil Gehrels Swift Observatory X-ray Telescope \citep{Burrows2005}, 
the constraints are somewhat relaxed. 
In this scenario, it is required that, 
after $\sim 10~{\rm days}$ from the first X-ray flare of the consecutive QPEs and after $\sim 70~{\rm days}$ from the sixth X-ray flare, 
the radio fluxes should be low ($\sim 1/6$ and $\sim 1/4$ of the detected radio flux, respectively). 
This scenario is illustrated in the upper panel of Fig.~\ref{fig:l_ev_obs}, where radio emission is produced by ejecta associated with the bright QPE flares in the 1st--6th, 10th, and 11th cycles. 
To ensure rapid decay of radio emission after the peak caused by gas ejection associated with the 6th cycle, we adjust a parameter for the wind profile to be steeper by setting it to $s=2.4$. 
Hence, the models with steeper decay for radio emission are required such as 
lower $n$, larger $p_{\rm e}$, and larger $s$.

As an alternative possibility, 
we can consider that the winds are cleared by collisions with ejecta from stellar interactions with the accretion disk. 
Consequently, the radio flare is produced solely by the ejecta generated during the first QPE flare. 
An example of this scenario is illustrated in the lower panel of Fig.~\ref{fig:l_ev_obs}, where the radio emission is generated exclusively by the ejecta associated with the first QPE flare (purple line). 
Note that a higher flux is likely required in this scenario, such as $m_{\rm ej}=10^{-4}~\Msun$ and $\epsilon_{\rm e}=0.4$, 
which is higher than those estimated from recent PIC simulations \citep[e.g.,][]{Jikei2026} and afterglow observations \citep[e.g.,][]{Panaitescu2001}.

These scenarios become implausible if bright QPEs existed a few cycles prior to the first flare identified by \citet{Guolo2024}. 
In either case, subsequent observations with dense sampling 
and/or multi-wavelength observations using the VLA, ALMA, and JWST 
are highly desirable to distinguish between these models.

\subsection{Preferred events}

We briefly discuss the reasons for the detection of radio emission in J0230+28. 
We hypothesize that a primary factor is the star's large orbital radius. 
Given the long recurrence time of this event, the orbital radius is estimated to be significantly larger than that of other events. 
Additionally, \citet{Yao2024} and \citet{Linial2025} suggest that the cross-section for stellar collisions with an accretion disk may be extended due to the expansion of the stellar envelope, which can reach up to the Hill radius. 
If this is indeed the case, 
the mass of the disk gas impacted by the stellar collision in this event is estimated to be more than a factor of $\sim 9$--$100$ times greater than in other events, based on the surface density calculated from the upper limit of luminosity from an accretion disk \citep{Haiman2009}.

On the other hand, for larger orbital radii, the orbital velocity tends to be lower, reducing the radio flux. However, the orbital velocity for observed QPEs ranges from $\sim 0.03$ to $0.1~c$ presumably to be bright in X-ray, with an impact on the flux of roughly up to one order of magnitude. 
If the flux is roughly proportional to $m_{\rm ej} v_{\rm ej}^2$, then the contribution from the disk mass is significant.

Overall, 
events with longer recurrence times have a greater likelihood of producing detectable radio counterparts.

\section{Discussions}

\label{sec:discussions}

In this section, we discuss the properties of the X-ray flares expected for the fiducial model in this paper, 
and their implications for the event, Swift~J0230+28. 

\begin{figure}
\begin{center}
\includegraphics[width=85mm]{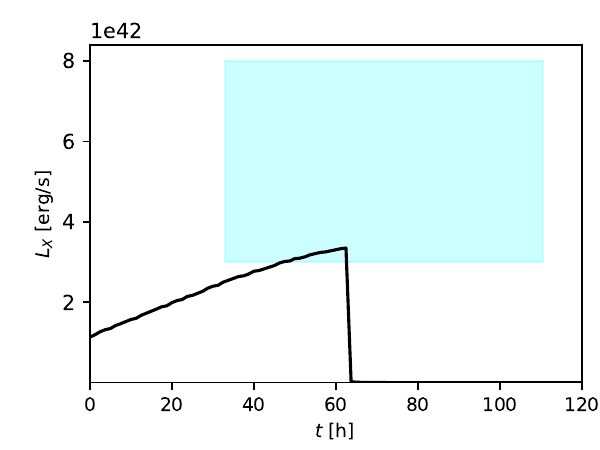}
\caption{
The light curves for the model with a highly inclined orbit adjusted to explain Swift~J0230+28. 
The cyan region along the X-axis represents the uncertainty of the rise phase, 
while that along the Y-axis represents the peak luminosity in the 1st-6th, 10th, and 11th flares. 
The roughness results from finite resolution and can be reduced by increasing the spatial grid resolution. 
}
\label{fig:lx_swift}
\end{center}
\end{figure}

\subsection{X-ray flares}

\subsubsection{Standard model}

We briefly discuss whether the fiducial parameters can account for the production of X-ray flares in Swift~J0230+28. 
In this subsection, 
we assume that 
the extended duration of this event can be explained 
either by elongated streams \citep{Yao2024,Linial2025} 
or by an extended disk \citep{ZhuZhaohuan2026}. 
The former may result from 
the expansion of the stellar envelope, while the latter may result from magnetic effects. 
The required length of the stream or the disk is 
\begin{align}
\label{eq:rlen}
R_{\rm len}\sim &v_{\rm Kep} t_{\rm dur}
\sim &2\times 10^{14}~{\rm cm}
\left(\frac{v_{\rm Kep}}{0.03~c}\right)
\left(\frac{t_{\rm dur}}{3~{\rm day}}\right)
\end{align}
along a direction of orbit.

The total energy of the flare is limited by the kinetic energy of the ejecta divided by the duration, given by 
\begin{align}
\label{eq:pkin}
P_{\rm kin}\sim &\frac{m_{\rm ej}v_{\rm ej}^2}{t_{\rm dur}}\nonumber\\
\sim &1\times 10^{43}~{\rm erg/s} 
\left(\frac{m_{\rm ej}}{4\times 10^{-4}~\Msun}\right)
\left(\frac{v_{\rm ej}}{0.06 c}\right)^{2}
\left(\frac{t_{\rm dur}}{3~{\rm day}}\right)^{-1},
\end{align}
where $t_{\rm dur}$ is the duration of the X-ray flares. The fiducial parameters can explain the observed X-ray luminosity of $L_{\rm QPE}\sim 6\times 10^{42}~{\rm erg/s}$, assuming efficient conversion of kinetic energy into radiation energy. 
In the case of cooling emission, 
the luminosity typically falls below $P_{\rm kin}$ due to energy loss from adiabatic expansion, unless the shocked gas rapidly becomes transparent after the shock front breakout. 
For breakout emission, the breakout timescale needs to be comparable to the observed duration to match the luminosity with $P_{\rm kin}$ for $t_{\rm dur}$. 
This could be achievable, as \citet{ZhuZhaohuan2026} found that $t_{\rm dur}\sim 0.2 P_{\rm QPE}$ by breakout emission with an extended disk structure.

The radiation temperature may be useful to constrain emission mechanism of X-ray flares. 
The black body temperature of the cooling emission is \citep{Linial2023,Tagawa2023_SC}
\begin{align}
\label{eq:tcool}
T_{\rm cool}
=&\left(\frac{L_{\rm QPE}\tau}{4\pi R_{\rm diff}^2 ca}\right)^{1/4}\nonumber\\
\sim& 2\times 10^4~{\rm K}
\left(\frac{L_{\rm QPE}}{6\times 10^{42}~{\rm erg/s}}\right)^{1/4}\nonumber\\
&\left(\frac{v_{\rm ej}}{0.06~c}\right)^{-3/4}
\left(\frac{t_{\rm dur}}{3~{\rm d}}\right)^{-1/2}, 
\end{align} 
where $\tau$ is the optical depth at the diffusion of cooling emission ($t=t_{\rm diff}$), 
$R_{\rm diff}\sim t_{\rm diff}v_{\rm ej}$ is the diffusion radius, 
$a$ is the radiation constant, and we assume $t_{\rm diff}\sim t_{\rm dur}$. 
Similarly, the black body temperature of the breakout emission is \citep{Nakar2010,Tagawa2023_breakout}
\begin{align}
T_{\rm break}= &(18 \rho_{\rm bf-sh} v_{\rm ej}^2 /7a)^{1/4}\nonumber\\
\sim & 10^6~{\rm K}
\left(\frac{m_{\rm ej}}{4\times 10^{-4}~\Msun}\right)^{1/4}\nonumber\\
&\left(\frac{R_{\rm BO}}{10^{13}~{\rm cm}}\right)^{-3/4}
\left(\frac{v_{\rm ej}}{0.06 c}\right)^{1/2}
\end{align} 
where $R_{\rm BO}$ is the characteristic size of the shocked gas at the breakout of a star from an accretion disk, and $\rho_{\rm bf-sh}$ is the density of gas before experiencing shocks that later produces an X-ray QPE flare. 
In the equations above, 
we adopt $R_{\rm BO}\sim (R_{\rm Hill}^2 R_{\rm len})^{1/3}\sim 10^{13}~{\rm cm}$. 
If thermal equilibrium is not achieved due to a lack of photons, 
the radiation temperature can be enhanced by a factor of ${\rm max}(1,\eta_0^2)$ in the breakout phase compared to the black body temperature \citep{Nakar2010}, where 
\begin{align}
\eta_0^2
\sim 0.5 
\left(\frac{v_{\rm ej}}{0.06 c}\right)^{15/2}
\left(\frac{m_{\rm ej}}{4\times 10^{-4}~\Msun}\right)^{-1/4}
\left(\frac{R_{\rm BO}}{10^{13}~{\rm cm}}\right)^{3/4}.
\end{align}
Note that in the cooling emission phase, $\eta_0$ is further reduced by the ratio of the diffusion time during the cooling phase over the diffusion time during the breakout phase. 
Since $\eta_0 \lesssim 1$, the radiation is expected to be in the equilibrium state. 
Hence, in the fiducial model, the observed radiation temperature ($\sim 1.4 \times 10^6~{\rm K}$) can be explained by emission in the breakout phase. 
On the other hand, it is difficult to explain the temperature by the cooling emission, since $m_{\rm ej}$ and $R_{\rm BO}$, which can affect the temperature, are constrained by luminosity and duration (Eqs.~\ref{eq:rlen}, \ref{eq:pkin}, and \ref{eq:tcool}).

\subsubsection{High inclination model}

Another possibility for explaining the extended duration is a highly inclined orbit. 
Even without an elongated stream or an extended disk, if the inclination angle of the star's orbit with respect to the disk is very low ($\theta_*={\rm asin}[(R_*/t_{\rm dur})/v_{\rm kep}]\sim 0.6^\circ$), 
a prolonged duration of $\sim 3~{\rm days}$ can be achieved with the fiducial parameter set. 
Such an inclination may occur when stars, initially embedded in an AGN disk, orbit in a retrograde direction (see \S~\ref{sec:formation}).

Since the duration and light curves for breakout emission in inclined cases are difficult to predict by analytic estimates, 
we use the method for calculating emission properties around a breakout phase developed in \citet{Tagawa2023}, 
and calculate the evolution of this X-ray flare for a highly inclined orbit. 
We set $\theta_*=1^\circ$, and use $r_{\rm Hill}$ as the stellar size 
($R_* = 2.6\times 10^{12}~{\rm cm}$). 
As a higher spatial resolution is required to predict light curves for small inclination angles, 
we set the spatial grid number to $10^4$. To reduce computational costs, we decreased the time grid number to $100$. 
Note that in low $\theta_*$ cases, $m_{\rm ej}$ is enhanced, which also increases the peak radio emission roughly by the same factor. This enhancement allows the reproduction of high radio luminosity with smaller values of plasma parameters ($\epsilon_{\rm e}$, $\epsilon_{\rm B}$). 

Fig.~\ref{fig:lx_swift} shows the time evolution of the X-ray luminosity for the fiducial model. 
In this setting, the luminosity changes by a factor of 2 over $\sim 1~{\rm day}$, which is comparable to the light curve observed in Swift~J0230+28 \citep{Guolo2024}. 
This long duration is achieved by adopting a large $R_*$ and a very low $\theta_*$. 
A large $R_*$ may result from the heating of the stellar envelope due to collisions \citep{Yao2024}, while a very low $\theta_*$ can be expected when considering the formation process of this system (\S~\ref{sec:formation}). 
The advantage of this method is its prediction of a slow rise and rapid decay, consistent with observations and not produced by other collision models. 
Note that in our model, the decay is too rapid because we do not account for shock cooling emission, consider a geometrically thin standard disk model, 
and 
assume a constant density for the accretion disk for simplicity \citep{Tagawa2023}. 
Further modeling will be required to accurately reproduce light curves during the decay phase.

\citet{Pasham2024_Swift} reported quiescent periods for X-ray flares, occurring before and after the active periods observed by \citet{Guolo2024}. 
Additionally, there are 
variations in the duration and flux of these flares. 
We propose that these variations can be explained by the evolution of the inclination angle due to the viscous evolution of possible warps or twists \citep[e.g.,][]{Natarajan1998,Wijers1999,Fragner2010} or by the precession of stellar orbits in warped disks \citep{Franchini+2023}. 
Due to these effects, the inclination angle could increase, decrease, or oscillate by several degrees. 
In this scenario, the duration is almost inversely proportional to $\theta_*$ according to our calculations, meaning that eruptions with short durations can be explained by an increase in $\theta_*$. 
The quiescent phases may be explained by either a considerably larger or lower $\theta_*$. 
When $\theta_*\approx 0^\circ$, the stellar orbit can be entirely embedded in the AGN disk, and breakout emission are not expected. Conversely, when $\theta_*\gtrsim 10^\circ$, the duration becomes $\lesssim 5~{\rm hr}$, suggesting that such events may be less pronounced and tend to be missed in the observations (see Fig.~4 in \citealt{Pasham2024_Swift}).

\subsection{Possible formation of system}

\label{sec:formation}

Here we propose that the formation process of Swift~J0230+28 may be distinct from that of other QPEs. 
Since many QPEs are associated with tidal disruption events, 
the accretion disks (or stars) in these events likely form due to the disruption of a star \citep{Miniutti2023_reQPE,Arcodia2024_QPE34,Nicholl2024}. 
However, the mass of the accretion disk within the orbit is significantly higher for Swift~J0230+28. If we estimate the gas mass within the orbit using the background (AGN) luminosity, the SMBH mass, a viscous parameter of $\alpha=0.02$, and $\eta_{\rm rad}=0.1$ (with $\eta_{\rm rad}=0.03$ for Swift~J0230+28 due to its low AGN luminosity) through the standard disk model \citep{Haiman2009}, 
the mass of the accretion disk within the orbit is estimated to be $\sim 0.03$--$7~\Msun$ for other events, whereas it is $\sim 30~\Msun$ for Swift~J0230+28. 
Although this mass is proportional to $\alpha^{-4/5}$ or $\alpha^{-1}$ depending on the regions and can be reduced by a factor of $5$ at $\alpha\sim 0.1$, it remains too high compared to the mass of usual stars. 
Hence, the accretion disks in Swift~J0230+28 are difficult to explain by mass loss from a star due to tidal disruption events or mass transfer from a tightly orbiting star.

We suggest that the accretion disk in this event formed through typical AGN disk-formation processes. 
In such cases, a tight retrograde star–disk configuration may also form. 
If a star is initially counter-rotating relative to the AGN disk, 
the timescale for the evolution of the semi-major axis becomes significantly shorter \citep{WangY2024_capture}. 
Assuming that the disk surface density is $\Sigma_{\rm disk}\sim 4\times 10^4~{\rm g~cm^{-3}}$ at $a\sim 10^3~r_{\rm g}$ (which is estimated using the AGN luminosity, SMBH mass, $\alpha=0.02$, and the location of the star in Swift~J0230+28, \citealt{Haiman2009}), 
and that the stellar mass is $m_*\sim 1~\Msun$ with a stellar radius of $R_*\sim 2.6\times 10^{12}~{\rm cm}$, 
the migration timescale is estimated as 
$t_{\rm mig}\sim f_{\rm mig} t_{\rm orb} (m_*/R_*^2)/\Sigma_{\rm disk} \sim {\rm yr}$, where $t_{\rm orb}$ is the orbital timescale of a star, set to be $\sim 2P_{\rm QPE}$, 
$f_{\rm mig}$ is the fraction representing the reduction timescale of the semi-major axis, and it is $f_{\rm mig}\sim 10^{-3}$ for a retrograde star \citep{WangY2024_capture}. 
Since this timescale is proportional to $a^{21/10}$--$a^{9/4}$ with a fixed opacity, 
the star can reduce its orbital radius by two to three orders of magnitude within the AGN lifetime of a few ${\rm Myr}$. 
During the migration, the star needs to survive ablation due to shocks from colliding with counter-rotating gas. 
Numerical simulations by \citet{Yao2024} suggest that the mass loss fraction per collision asymptotic to $\Delta m_*/m_*\sim f_{\rm ab} \times (p_{\rm ram}/p_*)$, where $p_{\rm ram}$ is the ram pressure, $p_*$ is the mean stellar pressure, and $f_{\rm ab}\sim 0.03$ is the fraction of mass lost per collision. We assume that this mass is lost over the thermal timescale evaluated at the stellar envelope where the internal pressure equals $p_{\rm ram}$ \citep{Yao2024}. 
Consequently, the ratio of the ablation timescale ($t_{\rm ab}\sim t_{\rm orb}m_*/\Delta m_*$) to the migration timescale is 
\begin{eqnarray}
\frac{t_{\rm ab}}{t_{\rm mig}}
\sim 100
\left(\frac{f_{\rm mig}}{10^{-3}}\right)^{-1}
\left(\frac{f_{\rm ab}}{0.03}\right)^{-1}\nonumber\\
\left(\frac{M_{\rm SMBH}/m_*}{10^{6.6}}\right)^{-1}
\left(\frac{H_{\rm disk}a/R_*^2}{10^5}\right),
\end{eqnarray}
which is much larger than $\sim 1$ at $a\gtrsim 100~r_g$, where stars producing QPEs typically reside. 
Thus, many retrograde stars that initially orbit at outer radii of $\lesssim 0.1$--$1~{\rm pc}$ can become tightly bound stars due to gas drag within $\sim 1$--$100~{\rm Myr}$ and can then be observed as QPEs. 

Note that the decay timescale of the eccentricity is $\sim 10$ times that of the semi-major axis, while the pumping timescale of the inclination is $\sim 10^3$ times that of the semi-major axis \citep{WangY2024_capture}. 
Hence, 
the inclination increases very slowly as the star experiences gas drag, 
which accounts for the very low inclination angle required to explain the long duration of this event in this scenario.

Moreover, the rate of significant migration for retrograde stars due to gas drag is estimated to be $\sim 10^{-4}~{\rm yr}^{-1}$ per galaxy \citep{WangY2024_TDE_AGN}, 
which is comparable to 
the occurrence rate of QPEs \citep{Arcodia2024}.

Overall, we propose that the origin of the accretion disk and the star's evolution in Swift~J0230+28 may differ from those in other cases, 
with the star migrating from larger scales due to gas torque. 
Note that both stream, extended disk, and high-inclination models can explain the properties of the observed radio emission and X-ray QPEs. To distinguish these models, detailed observations of similar systems are useful.

\section{Conclusions}

\label{sec:conclusions}

In this study, 
we investigate the properties of synchrotron emission from shocks associated with collisions between the gas ejected by a colliding star with the accretion disk 
and the winds emitted from the accretion disk (or from previous QPE episodes). 
We also discuss whether this process can explain the detection and non-detection of radio emission observed in Swift~J0230+28. 
Our findings are as follows.

\begin{enumerate}

\item 
The synchrotron emission can explain both the detection and non-detection of radio flares 
through the collision model. 
To reproduce the observed flux, high efficiencies for the conversion of shock kinetic energy to radiation energy are required. 

\item 
Due to synchrotron self-absorption, 
the evolution of the radio flux is expected to be gradual, with a rise timescale of several tens of days considering the properties of radio emission in Swift~J0230+28. 
This constrains the active phases during which the ejecta are produced, subsequently  colliding with the winds and generating synchrotron emission.

\item 
The detection and non-detection of radio fluxes can be explained, 
if ejecta are launched during cycles with bright QPEs, specifically from the first to the sixth cycles, and not during the less bright cycles, such as the seventh to ninth cycles as observed by \citet{Guolo2024}. 
Another possibility is that synchrotron emission is generated solely by ejecta from the first cycle, with the shocks clearing the winds. 
Dense sampling and multi-wavelength observations using the VLA, ALMA, and JWST 
could provide valuable constraints on these scenarios, as well as on the intrinsic parameters of these systems.

\item 
Radio counterparts are easier to detect 
for events with longer recurrence times, 
since higher ejecta masses lead to brighter radio emission. 

\item 
The X-ray emission of QPE flares in Swift~J0230+28 is better explained by breakout emission rather than cooling emission, aligning with the observed radiation temperature. 
The long duration might result from an elongated stream, an extended disk structure, or a highly inclined orbit. 
The slow rise and rapid decay of the light curves can be reproduced by breakout emission from such an inclined orbit. 
This inclination may occur if a retrograde star migrates from the outer regions of an AGN disk, differing from many other events where accretion disks likely form due to the tidal disruption of a star.

\end{enumerate}

\section*{acknowledgments}


We thank Clement Bonnerot for the fruitful discussions and for suggesting the intriguing possibility that shocks occur between ejecta produced during different QPE episodes. 
H.T. is supported by the National Science and Technology Major Project of China (No. 2024ZD1100601) and 
the National Key R$\&$D Program of China (grant No.2024YFC2207700). 
S.S.K. acknowledge the support by JSPS KAKENHI Nos. JP23H04899, JP26K00733, JP26K00696, and  the Tohoku Initiative for Fostering Global Researchers for Interdisciplinary Sciences (TI-FRIS) of MEXT’s Strategic Professional Development Program for Young Researchers.

\section*{Data availability}

The data underlying this article are available in the article.

\appendix

\bibliographystyle{aasjournal}
\bibliography{agn_bhm}

@ARTICLE{Chakraborty2026_warp,
       author = {{Chakraborty}, Joheen and {Mummery}, Andrew and {Quataert}, Eliot and {Arcodia}, Riccardo and {Linial}, Itai and {Kara}, Erin},
        title = "{Quasi-periodic eruption spectral-timing: EMRIs crossing warped accretion disks}",
      journal = {arXiv e-prints},
         year = 2026,
        month = oct,
          eid = {arXiv:2610.10110},
        pages = {arXiv:2610.10110},
          doi = {10.48550/arXiv.2610.10110},
archivePrefix = {arXiv},
       eprint = {2610.10110},
 primaryClass = {astro-ph.HE},
       adsurl = {https://ui.adsabs.harvard.edu/abs/2026arXiv261010110C}
}

@ARTICLE{ZhangZirui2026,
       author = {{Zhang}, Zirui and {Zhou}, Cong and {Pan}, Zhen},
        title = "{Secular Evolution of Quasiperiodic Eruptions: eRO-QPE2}",
      journal = {arXiv e-prints},
         year = 2026,
        month = oct,
          eid = {arXiv:2610.21844},
        pages = {arXiv:2610.21844},
          doi = {10.48550/arXiv.2610.21844},
archivePrefix = {arXiv},
       eprint = {2610.21844},
 primaryClass = {astro-ph.HE},
       adsurl = {https://ui.adsabs.harvard.edu/abs/2026arXiv261021844Z}
}

@ARTICLE{ChenBoAn2026,
       author = {{Chen}, Bo-An and {You}, Bei and {Miniutti}, Giovanni and {Jiang}, Ning and {Pan}, Zhen and {Yang}, Tao and {Fan}, Xi-Long and {Liao}, Kai and {Ding}, Xu-Heng and {Zhu}, Zong-Hong and {Yang}, Shuai-Kang and {Xu}, Sai-En and {He}, Han and {Fan}, Xiao},
        title = "{QPEs from Warped Disk Collisions with EMRIs: Brightness-Recurrence Diagram and Gravitational-Wave Follow-up}",
      journal = {arXiv e-prints},
         year = 2026,
        month = may,
          eid = {arXiv:2605.24905},
        pages = {arXiv:2605.24905},
          doi = {10.48550/arXiv.2605.24905},
archivePrefix = {arXiv},
       eprint = {2605.24905},
 primaryClass = {astro-ph.HE},
       adsurl = {https://ui.adsabs.harvard.edu/abs/2026arXiv260524905C}
}

@ARTICLE{Williamson2020,
       author = {{Williamson}, David and {H{\"o}nig}, Sebastian and {Venanzi}, Marta},
        title = "{Radiation Hydrodynamics Models of Active Galactic Nuclei: Beyond the Central Parsec}",
      journal = {\apj},
         year = 2020,
        month = jul,
       volume = {897},
       number = {1},
          eid = {26},
        pages = {26},
          doi = {10.3847/1538-4357/ab989e},
archivePrefix = {arXiv},
       eprint = {2006.00918},
 primaryClass = {astro-ph.GA},
       adsurl = {https://ui.adsabs.harvard.edu/abs/2020ApJ...897...26W}
}

@ARTICLE{Kudoh2023,
       author = {{Kudoh}, Yuki and {Wada}, Keiichi and {Kawakatu}, Nozomu and {Nomura}, Mariko},
        title = "{Multiphase Gas Nature in the Sub-parsec Region of the Active Galactic Nuclei. I. Dynamical Structures of Dusty and Dust-free Outflow}",
      journal = {\apj},
         year = 2023,
        month = jun,
       volume = {950},
       number = {1},
          eid = {72},
        pages = {72},
          doi = {10.3847/1538-4357/accc2b},
archivePrefix = {arXiv},
       eprint = {2304.05950},
 primaryClass = {astro-ph.GA},
       adsurl = {https://ui.adsabs.harvard.edu/abs/2023ApJ...950...72K}
}

@ARTICLE{Wada2012,
       author = {{Wada}, Keiichi},
        title = "{Radiation-driven Fountain and Origin of Torus around Active Galactic Nuclei}",
      journal = {\apj},
         year = 2012,
        month = oct,
       volume = {758},
       number = {1},
          eid = {66},
        pages = {66},
          doi = {10.1088/0004-637X/758/1/66},
archivePrefix = {arXiv},
       eprint = {1208.5272},
 primaryClass = {astro-ph.GA},
       adsurl = {https://ui.adsabs.harvard.edu/abs/2012ApJ...758...66W}
}

@ARTICLE{Lyu2022,
       author = {{Lyu}, Jianwei and {Rieke}, George},
        title = "{Infrared Spectral Energy Distribution and Variability of Active Galactic Nuclei: Clues to the Structure of Circumnuclear Material}",
      journal = {Universe},
         year = 2022,
        month = may,
       volume = {8},
       number = {6},
          eid = {304},
        pages = {304},
          doi = {10.3390/universe8060304},
archivePrefix = {arXiv},
       eprint = {2205.14172},
 primaryClass = {astro-ph.GA},
       adsurl = {https://ui.adsabs.harvard.edu/abs/2022Univ....8..304L}
}

@ARTICLE{Jikei2026,
       author = {{Jikei}, Taiki and {Groselj}, Daniel and {Sironi}, Lorenzo},
        title = "{Dependence of Particle Acceleration Efficiency on Shock Velocity in Weakly Magnetized Electron-Ion Shocks}",
      journal = {arXiv e-prints},
         year = 2026,
        month = jul,
          eid = {arXiv:2607.05778},
        pages = {arXiv:2607.05778},
          doi = {10.48550/arXiv.2607.05778},
archivePrefix = {arXiv},
       eprint = {2607.05778},
 primaryClass = {astro-ph.HE},
       adsurl = {https://ui.adsabs.harvard.edu/abs/2026arXiv260705778J}
}

@ARTICLE{Matsuoka2020,
       author = {{Matsuoka}, Tomoki and {Maeda}, Keiichi},
        title = "{Radio Emission from Ultra-stripped Supernovae as Diagnostics for Properties of the Remnant Double Neutron Star Binaries}",
      journal = {\apj},
         year = 2020,
        month = aug,
       volume = {898},
       number = {2},
          eid = {158},
        pages = {158},
          doi = {10.3847/1538-4357/ab9c1b},
archivePrefix = {arXiv},
       eprint = {2006.06689},
 primaryClass = {astro-ph.HE},
       adsurl = {https://ui.adsabs.harvard.edu/abs/2020ApJ...898..158M}
}

@ARTICLE{Suzuguchi2026,
       author = {{Suzuguchi}, Tomoya and {Matsumoto}, Tatsuya},
        title = "{Quasi-periodic eruptions as a probe of accretion disc in tidal disruption events}",
      journal = {\mnras},
         year = 2026,
        month = feb,
       volume = {545},
       number = {4},
          eid = {staf2243},
        pages = {staf2243},
          doi = {10.1093/mnras/staf2243},
archivePrefix = {arXiv},
       eprint = {2509.01663},
 primaryClass = {astro-ph.HE},
       adsurl = {https://ui.adsabs.harvard.edu/abs/2026MNRAS.545f2243S}
}

@INPROCEEDINGS{DiFrancesco2019,
       author = {{Di Francesco}, James and {Chalmers}, Dean and {Denman}, Nolan and {Fissel}, Laura and {Friesen}, Rachel and {Gaensler}, Bryan and {Hlavacek-Larrondo}, Julie and {Kirk}, Helen and {Matthews}, Brenda and {O'Dea}, Christopher and {Robishaw}, Tim and {Rosolowsky}, Erik and {Rupen}, Michael and {Sadavoy}, Sarah and {Sa-Harb}, Samar and {Sivakoff}, Greg and {Tahani}, Mehrnoosh and {van der Marel}, Nienke and {White}, Jacob and {Wilson}, Christine},
        title = "{The Next Generation Very Large Array}",
    booktitle = {Canadian Long Range Plan for Astronomy and Astrophysics White Papers},
         year = 2019,
       volume = {2020},
        month = oct,
          eid = {32},
        pages = {32},
          doi = {10.5281/zenodo.3765763},
archivePrefix = {arXiv},
       eprint = {1911.01517},
 primaryClass = {astro-ph.IM},
       adsurl = {https://ui.adsabs.harvard.edu/abs/2019clrp.2020...32D}
}

@ARTICLE{Rigby2023,
       author = {{Rigby}, Jane and {Perrin}, Marshall and {McElwain}, Michael and {Kimble}, Randy and {Friedman}, Scott and {Lallo}, Matt and {Doyon}, Ren{\'e} and {Feinberg}, Lee and {Ferruit}, Pierre and {Glasse}, Alistair and {Rieke}, Marcia and {Rieke}, George and {Wright}, Gillian and {Willott}, Chris and {Colon}, Knicole and {Milam}, Stefanie and {Neff}, Susan and {Stark}, Christopher and {Valenti}, Jeff and {Abell}, Jim and {Abney}, Faith and {Abul-Huda}, Yasin and {Acton}, D. Scott and {Adams}, Evan and {Adler}, David and {Aguilar}, Jonathan and {Ahmed}, Nasif and {Albert}, Lo{\"\i}c and {Alberts}, Stacey and {Aldridge}, David and {Allen}, Marsha and {Altenburg}, Martin and {{\'A}lvarez-M{\'a}rquez}, Javier and {Alves de Oliveira}, Catarina and {Andersen}, Greg and {Anderson}, Harry and {Anderson}, Sara and {Argyriou}, Ioannis and {Armstrong}, Amber and {Arribas}, Santiago and {Artigau}, Etienne and {Arvai}, Amanda and {Atkinson}, Charles and {Bacon}, Gregory and {Bair}, Thomas and {Banks}, Kimberly and {Barrientes}, Jaclyn and {Barringer}, Bruce and {Bartosik}, Peter and {Bast}, William and {Baudoz}, Pierre and {Beatty}, Thomas and {Bechtold}, Katie and {Beck}, Tracy and {Bergeron}, Eddie and {Bergkoetter}, Matthew and {Bhatawdekar}, Rachana and {Birkmann}, Stephan and {Blazek}, Ronald and {Blome}, Claire and {Boccaletti}, Anthony and {B{\"o}ker}, Torsten and {Boia}, John and {Bonaventura}, Nina and {Bond}, Nicholas and {Bosley}, Kari and {Boucarut}, Ray and {Bourque}, Matthew and {Bouwman}, Jeroen and {Bower}, Gary and {Bowers}, Charles and {Boyer}, Martha and {Bradley}, Larry and {Brady}, Greg and {Braun}, Hannah and {Breda}, David and {Bresnahan}, Pamela and {Bright}, Stacey and {Britt}, Christopher and {Bromenschenkel}, Asa and {Brooks}, Brian and {Brooks}, Keira and {Brown}, Bob and {Brown}, Matthew and {Brown}, Patricia and {Bunker}, Andy and {Burger}, Matthew and {Bushouse}, Howard and {Cale}, Steven and {Cameron}, Alex and {Cameron}, Peter and {Canipe}, Alicia and {Caplinger}, James and {Caputo}, Francis and {Cara}, Mihai and {Carey}, Larkin and {Carniani}, Stefano and {Carrasquilla}, Maria and {Carruthers}, Margaret and {Case}, Michael and {Catherine}, Riggs and {Chance}, Don and {Chapman}, George and {Charlot}, St{\'e}phane and {Charlow}, Brian and {Chayer}, Pierre and {Chen}, Bin and {Cherinka}, Brian and {Chichester}, Sarah and {Chilton}, Zack and {Chonis}, Taylor and {Clampin}, Mark and {Clark}, Charles and {Clark}, Kerry and {Coe}, Dan and {Coleman}, Benee and {Comber}, Brian and {Comeau}, Tom and {Connolly}, Dennis and {Cooper}, James and {Cooper}, Rachel and {Coppock}, Eric and {Correnti}, Matteo and {Cossou}, Christophe and {Coulais}, Alain and {Coyle}, Laura and {Cracraft}, Misty and {Curti}, Mirko and {Cuturic}, Steven and {Davis}, Katherine and {Davis}, Michael and {Dean}, Bruce and {DeLisa}, Amy and {deMeester}, Wim and {Dencheva}, Nadia and {Dencheva}, Nadezhda and {DePasquale}, Joseph and {Deschenes}, Jeremy and {Hunor Detre}, {\"O}rs and {Diaz}, Rosa and {Dicken}, Dan and {DiFelice}, Audrey and {Dillman}, Matthew and {Dixon}, William and {Doggett}, Jesse and {Donaldson}, Tom and {Douglas}, Rob and {DuPrie}, Kimberly and {Dupuis}, Jean and {Durning}, John and {Easmin}, Nilufar and {Eck}, Weston and {Edeani}, Chinwe and {Egami}, Eiichi and {Ehrenwinkler}, Ralf and {Eisenhamer}, Jonathan and {Eisenhower}, Michael and {Elie}, Michelle and {Elliott}, James and {Elliott}, Kyle and {Ellis}, Tracy and {Engesser}, Michael and {Espinoza}, Nestor and {Etienne}, Odessa and {Etxaluze}, Mireya and {Falini}, Patrick and {Feeney}, Matthew and {Ferry}, Malcolm and {Filippazzo}, Joseph and {Fincham}, Brian and {Fix}, Mees and {Flagey}, Nicolas and {Florian}, Michael and {Flynn}, Jim and {Fontanella}, Erin and {Ford}, Terrance and {Forshay}, Peter and {Fox}, Ori and {Franz}, David and {Fu}, Henry and {Fullerton}, Alexander and {Galkin}, Sergey and {Galyer}, Anthony and {Garc{\'\i}a Mar{\'\i}n}, Macarena and {Gardner}, Jonathan P. and {Gardner}, Lisa and {Garland}, Dennis and {Garrett}, Bruce and {Gasman}, Danny and {Gaspar}, Andras and {Gaudreau}, Daniel and {Gauthier}, Peter and {Geers}, Vincent and {Geithner}, Paul and {Gennaro}, Mario and {Giardino}, Giovanna and {Girard}, Julien and {Giuliano}, Mark and {Glassmire}, Kirk and {Glauser}, Adrian},
        title = "{The Science Performance of JWST as Characterized in Commissioning}",
      journal = {\pasp},
         year = 2023,
        month = apr,
       volume = {135},
       number = {1046},
          eid = {048001},
        pages = {048001},
          doi = {10.1088/1538-3873/acb293},
archivePrefix = {arXiv},
       eprint = {2207.05632},
 primaryClass = {astro-ph.IM},
       adsurl = {https://ui.adsabs.harvard.edu/abs/2023PASP..135d8001R}
}

@ARTICLE{Chakraborty2026_outflow,
       author = {{Chakraborty}, Joheen and {Kara}, Erin and {Lu}, Wenbin and {Metzger}, Brian D. and {Kosec}, Peter and {Arcodia}, Riccardo and {Linial}, Itai and {Aspegren}, Olivia and {Behar}, Ehud and {Bhattacharyya}, Sudip and {Giustini}, Margherita and {Hernandez-Garcia}, Lorena and {Kasen}, Daniel and {Miniutti}, Giovanni and {Paerels}, Frits and {Quintin}, Erwan and {Ricci}, Claudio and {Rogantini}, Daniele and {Sanchez-Saez}, Paula and {Zaidouni}, Fatima},
        title = "{Relativistic outflows power a quasi-periodic eruption: constraints on energetics, mass loss, and emission mechanisms}",
      journal = {arXiv e-prints},
         year = 2026,
        month = aug,
          eid = {arXiv:2608.28507},
        pages = {arXiv:2608.28507},
          doi = {10.48550/arXiv.2608.28507},
archivePrefix = {arXiv},
       eprint = {2608.28507},
 primaryClass = {astro-ph.HE},
       adsurl = {https://ui.adsabs.harvard.edu/abs/2026arXiv260828507C}
}

@ARTICLE{WuTao2026,
       author = {{Wu}, Tao and {Shu}, Xinwen and {Sun}, Luming and {Jiang}, Ning and {Zhu}, Jiazheng and {Zhang}, Wenjie},
        title = "{Abnormal Nitrogen Abundance in the X-ray Spectrum of Quasi-periodically Erupting Source AT2019wzc}",
      journal = {arXiv e-prints},
         year = 2026,
        month = aug,
          eid = {arXiv:2608.01931},
        pages = {arXiv:2608.01931},
          doi = {10.48550/arXiv.2608.01931},
archivePrefix = {arXiv},
       eprint = {2608.01931},
 primaryClass = {astro-ph.HE},
       adsurl = {https://ui.adsabs.harvard.edu/abs/2026arXiv260801931W}
}

@ARTICLE{Shepherd2026ApJ,
       author = {{Shepherd}, Margaret and {French}, K. Decker and {Hinkle}, Jason T. and {Ferdinand} and {Mondal}, Samaresh and {Moon}, Yashasvi and {Verrico}, Margaret E.},
        title = "{The Delay Time Distribution of Quasiperiodic Eruptions}",
      journal = {\apjl},
         year = 2026,
        month = sep,
       volume = {1008},
       number = {2},
          eid = {L51},
        pages = {L51},
          doi = {10.3847/2041-8213/ae9c30},
archivePrefix = {arXiv},
       eprint = {2608.12261},
 primaryClass = {astro-ph.GA},
       adsurl = {https://ui.adsabs.harvard.edu/abs/2026ApJ..1008L..51S}
}

@ARTICLE{ZhuZhaohuan2026,
       author = {{Zhu}, Zhaohuan and {Huang}, Xiaoshan and {Jiang}, Yan-Fei and {Huang}, Shunquan},
        title = "{Unlocking the QPE Mystery: Star-Disk Collisions in Realistic AGN Disks}",
      journal = {arXiv e-prints},
         year = 2026,
        month = sep,
          eid = {arXiv:2609.03011},
        pages = {arXiv:2609.03011},
          doi = {10.48550/arXiv.2609.03011},
archivePrefix = {arXiv},
       eprint = {2609.03011},
 primaryClass = {astro-ph.HE},
       adsurl = {https://ui.adsabs.harvard.edu/abs/2026arXiv260903011Z}
}

@ARTICLE{Yao2026,
       author = {{Yao}, Philippe Z. and {Quataert}, Eliot and {Jiang}, Yan-Fei and {Linial}, Itai},
        title = "{Star-Disk Collisions II: Debris Stream Dynamics and Implications for QPEs and Other Transients Near SMBHs}",
      journal = {arXiv e-prints},
         year = 2026,
        month = jul,
          eid = {arXiv:2607.08823},
        pages = {arXiv:2607.08823},
          doi = {10.48550/arXiv.2607.08823},
archivePrefix = {arXiv},
       eprint = {2607.08823},
 primaryClass = {astro-ph.HE},
       adsurl = {https://ui.adsabs.harvard.edu/abs/2026arXiv260708823Y}
}

@ARTICLE{Jankovivc2026b,
       author = {{Jankovi{\v{c}}}, Taj and {Karpov}, Sergey and {Zaja{\v{c}}ek}, Michal and {Karas}, Vladim{\'\i}r and {{\'S}niegowska}, Marzena},
        title = "{Radiation-hydrodynamics of star-disc collisions: From system parameters to outflows and lightcurves}",
      journal = {arXiv e-prints},
         year = 2026,
        month = jul,
          eid = {arXiv:2607.05508},
        pages = {arXiv:2607.05508},
          doi = {10.48550/arXiv.2607.05508},
archivePrefix = {arXiv},
       eprint = {2607.05508},
 primaryClass = {astro-ph.HE},
       adsurl = {https://ui.adsabs.harvard.edu/abs/2026arXiv260705508J}
}

@ARTICLE{Maeda2013,
       author = {{Maeda}, Keiichi},
        title = "{Young Supernovae as Experimental Sites for Studying the Electron Acceleration Mechanism}",
      journal = {\apjl},
         year = 2013,
        month = jan,
       volume = {762},
       number = {2},
          eid = {L24},
        pages = {L24},
          doi = {10.1088/2041-8205/762/2/L24},
archivePrefix = {arXiv},
       eprint = {1211.5835},
 primaryClass = {astro-ph.HE},
       adsurl = {https://ui.adsabs.harvard.edu/abs/2013ApJ...762L..24M}
}

@ARTICLE{Caprioli2014,
       author = {{Caprioli}, D. and {Spitkovsky}, A.},
        title = "{Simulations of Ion Acceleration at Non-relativistic Shocks. I. Acceleration Efficiency}",
      journal = {\apj},
         year = 2014,
        month = mar,
       volume = {783},
       number = {2},
          eid = {91},
        pages = {91},
          doi = {10.1088/0004-637X/783/2/91},
archivePrefix = {arXiv},
       eprint = {1310.2943},
 primaryClass = {astro-ph.HE},
       adsurl = {https://ui.adsabs.harvard.edu/abs/2014ApJ...783...91C}
}

@ARTICLE{Tagawa2023_breakout,
       author = {{Tagawa}, Hiromichi and {Kimura}, Shigeo S. and {Haiman}, Zolt{\'a}n and {Perna}, Rosalba and {Bartos}, Imre},
        title = "{Observable Signature of Merging Stellar-mass Black Holes in Active Galactic Nuclei}",
      journal = {\apj},
         year = 2023,
        month = jun,
       volume = {950},
       number = {1},
          eid = {13},
        pages = {13},
          doi = {10.3847/1538-4357/acc4bb},
archivePrefix = {arXiv},
       eprint = {2301.07111},
 primaryClass = {astro-ph.HE},
       adsurl = {https://ui.adsabs.harvard.edu/abs/2023ApJ...950...13T}
}

@ARTICLE{Czerny2004,
       author = {{Czerny}, B. and {R{\'o}za{\'n}ska}, A. and {Kuraszkiewicz}, J.},
        title = "{Constraints for the accretion disk evaporation rate in AGN from the existence of the Broad Line Region}",
      journal = {\aap},
         year = 2004,
        month = dec,
       volume = {428},
        pages = {39-49},
          doi = {10.1051/0004-6361:20040487},
archivePrefix = {arXiv},
       eprint = {astro-ph/0403507},
 primaryClass = {astro-ph},
       adsurl = {https://ui.adsabs.harvard.edu/abs/2004A&A...428...39C}
}

@ARTICLE{Evans2023,
       author = {{Evans}, P.~A. and {Nixon}, C.~J. and {Campana}, S. and {Charalampopoulos}, P. and {Perley}, D.~A. and {Breeveld}, A.~A. and {Page}, K.~L. and {Oates}, S.~R. and {Eyles-Ferris}, R.~A.~J. and {Malesani}, D.~B. and {Izzo}, L. and {Goad}, M.~R. and {O'Brien}, P.~T. and {Osborne}, J.~P. and {Sbarufatti}, B.},
        title = "{Monthly quasi-periodic eruptions from repeated stellar disruption by a massive black hole}",
      journal = {Nature Astronomy},
         year = 2023,
        month = nov,
       volume = {7},
        pages = {1368-1375},
          doi = {10.1038/s41550-023-02073-y},
archivePrefix = {arXiv},
       eprint = {2309.02500},
 primaryClass = {astro-ph.HE},
       adsurl = {https://ui.adsabs.harvard.edu/abs/2023NatAs...7.1368E}
}

@ARTICLE{YangLei2022,
       author = {{Yang}, Lei and {Shu}, Xinwen and {Zhang}, Fabao and {Chandola}, Yogesh and {Liu}, Daizhong and {Liu}, Yi and {Gu}, Minfeng and {Giustini}, Margherita and {Jiang}, Ning and {Li}, Ya-Ping and {Li}, Di and {Elbaz}, David and {Juneau}, Stephanie and {Pannella}, Maurilio and {Sun}, Luming and {Tang}, Ningyu and {Wang}, Tinggui and {Zhou}, Hongyan},
        title = "{Compact and Variable Radio Emission from an Active Galaxy with Supersoft X-Ray Emission}",
      journal = {\apj},
         year = 2022,
        month = aug,
       volume = {935},
       number = {2},
          eid = {115},
        pages = {115},
          doi = {10.3847/1538-4357/ac80ba},
archivePrefix = {arXiv},
       eprint = {2207.06585},
 primaryClass = {astro-ph.GA},
       adsurl = {https://ui.adsabs.harvard.edu/abs/2022ApJ...935..115Y}
}

@ARTICLE{Giustini2024,
       author = {{Giustini}, Margherita and {Miniutti}, Giovanni and {Arcodia}, Riccardo and {Goodwin}, Adelle and {Alexander}, Kate D. and {Chakraborty}, Joheen and {Buchner}, Johannes and {Kosec}, Peter and {Saxton}, Richard and {Bonetti}, Matteo and {Franchini}, Alessia and {Ryu}, Taeho and {Shu}, Xinwen and {Kara}, Erin and {Ponti}, Gabriele and {Quintin}, Erwan and {Vincentelli}, Federico and {Webb}, Natalie and {Kajava}, Jari and {von Fellenberg}, Sebastiano D.},
        title = "{Fragments of harmony amid apparent chaos: A closer look at the X-ray quasi-periodic eruptions of the galaxy RX J1301.9+2747}",
      journal = {\aap},
         year = 2024,
        month = dec,
       volume = {692},
          eid = {A15},
        pages = {A15},
          doi = {10.1051/0004-6361/202450861},
archivePrefix = {arXiv},
       eprint = {2409.01938},
 primaryClass = {astro-ph.HE},
       adsurl = {https://ui.adsabs.harvard.edu/abs/2024A&A...692A..15G}
}

@ARTICLE{Shu2017,
       author = {{Shu}, X.~W. and {Wang}, T.~G. and {Jiang}, N. and {Wang}, J.~X. and {Sun}, L.~M. and {Zhou}, H.~Y.},
        title = "{Central Engine and Host Galaxy of RXJ 1301.9+2747: A Multiwavelength View of a Low-mass Black Hole Active Galactic Nuclei with Ultra-soft X-Ray Emission}",
      journal = {\apj},
         year = 2017,
        month = mar,
       volume = {837},
       number = {1},
          eid = {3},
        pages = {3},
          doi = {10.3847/1538-4357/aa5eb3},
archivePrefix = {arXiv},
       eprint = {1704.02886},
 primaryClass = {astro-ph.HE},
       adsurl = {https://ui.adsabs.harvard.edu/abs/2017ApJ...837....3S}
}

@ARTICLE{Goodwin2025,
       author = {{Goodwin}, Adelle Jane and {Arcodia}, Riccardo and {Miniutti}, Giovanni and {Miller-Jones}, James Carl Alexander and {van Velzen}, Sjoert},
        title = "{The radio properties of quasi-periodic X-ray eruption sources}",
      journal = {Publications of the Astronomical Society of Australia},
         year = 2025,
        month = sep,
       volume = {42},
          eid = {e130},
        pages = {e130},
          doi = {10.1017/pasa.2025.10083},
archivePrefix = {arXiv},
       eprint = {2506.14417},
 primaryClass = {astro-ph.HE},
       adsurl = {https://ui.adsabs.harvard.edu/abs/2025PASA...42..130G}
}

@ARTICLE{Linial2026,
       author = {{Linial}, Itai and {Metzger}, Brian D. and {Beloborodov}, Andrei M.},
        title = "{Delayed Radio Flares in Tidal Disruption Events from Star-Disk Collision Outflows}",
      journal = {arXiv e-prints},
         year = 2026,
        month = may,
          eid = {arXiv:2605.28928},
        pages = {arXiv:2605.28928},
          doi = {10.48550/arXiv.2605.28928},
archivePrefix = {arXiv},
       eprint = {2605.28928},
 primaryClass = {astro-ph.HE},
       adsurl = {https://ui.adsabs.harvard.edu/abs/2026arXiv260528928L}
}

@ARTICLE{Fransson1998,
       author = {{Fransson}, Claes and {Bj{\"o}rnsson}, Claes-Ingvar},
        title = "{Radio Emission and Particle Acceleration in SN 1993J}",
      journal = {\apj},
         year = 1998,
        month = dec,
       volume = {509},
       number = {2},
        pages = {861-878},
          doi = {10.1086/306531},
archivePrefix = {arXiv},
       eprint = {astro-ph/9807030},
 primaryClass = {astro-ph},
       adsurl = {https://ui.adsabs.harvard.edu/abs/1998ApJ...509..861F}
}

@ARTICLE{Meszaros1997,
       author = {{M{\'e}sz{\'a}ros}, P. and {Rees}, M.~J.},
        title = "{Optical and Long-Wavelength Afterglow from Gamma-Ray Bursts}",
      journal = {\apj},
         year = 1997,
        month = feb,
       volume = {476},
       number = {1},
        pages = {232-237},
          doi = {10.1086/303625},
archivePrefix = {arXiv},
       eprint = {astro-ph/9606043},
 primaryClass = {astro-ph},
       adsurl = {https://ui.adsabs.harvard.edu/abs/1997ApJ...476..232M}
}

@ARTICLE{Pelle2026,
       author = {{Pelle}, Joaquin and {Kawaguchi}, Kyohei and {Shibata}, Masaru and {Tsz-Lok Lam}, Alan},
        title = "{Accretion-powered flares from black hole-disk collisions in galactic nuclei}",
      journal = {arXiv e-prints},
         year = 2026,
        month = apr,
          eid = {arXiv:2604.07443},
        pages = {arXiv:2604.07443},
          doi = {10.48550/arXiv.2604.07443},
archivePrefix = {arXiv},
       eprint = {2604.07443},
 primaryClass = {astro-ph.HE},
       adsurl = {https://ui.adsabs.harvard.edu/abs/2026arXiv260407443P}
}

@ARTICLE{Mummery2025,
       author = {{Mummery}, Andrew},
        title = "{Collisions with tidal disruption event disks: implications for quasi-periodic X-ray eruptions}",
      journal = {arXiv e-prints},
         year = 2025,
        month = apr,
          eid = {arXiv:2504.21456},
        pages = {arXiv:2504.21456},
          doi = {10.48550/arXiv.2504.21456},
archivePrefix = {arXiv},
       eprint = {2504.21456},
 primaryClass = {astro-ph.HE},
       adsurl = {https://ui.adsabs.harvard.edu/abs/2025arXiv250421456M}
}

@ARTICLE{Guolo2025_disk,
       author = {{Guolo}, M. and {Mummery}, A. and {Ingram}, A. and {Nicholl}, M. and {Gezari}, S. and {Nathan}, E.},
        title = "{A Time-dependent Solution for GSN 069 Disk Evolution and the Nature of Long-lived Tidal Disruption Events}",
      journal = {\apj},
         year = 2025,
        month = oct,
       volume = {992},
       number = {1},
          eid = {114},
        pages = {114},
          doi = {10.3847/1538-4357/ae039e},
archivePrefix = {arXiv},
       eprint = {2504.20148},
 primaryClass = {astro-ph.HE},
       adsurl = {https://ui.adsabs.harvard.edu/abs/2025ApJ...992..114G}
}

@ARTICLE{Mondek2026,
       author = {{Mondek}, Martin and {Zaja{\v{c}}ek}, Michal and {Best}, Henry and {Jankovi{\v{c}}}, Taj and {Karas}, Vladim{\'\i}r and {Kurf{\"u}rst}, Petr},
        title = "{Evolution of Quasi-Periodic Eruptions in the post-TDE Accretion Disk Perturbed by an Orbiting Star}",
      journal = {arXiv e-prints},
         year = 2026,
        month = mar,
          eid = {arXiv:2603.28380},
        pages = {arXiv:2603.28380},
          doi = {10.48550/arXiv.2603.28380},
archivePrefix = {arXiv},
       eprint = {2603.28380},
 primaryClass = {astro-ph.HE},
       adsurl = {https://ui.adsabs.harvard.edu/abs/2026arXiv260328380M}
}

@ARTICLE{Bykov2025,
       author = {{Bykov}, S.~D. and {Gilfanov}, M.~R. and {Sunyaev}, R.~A. and {Medvedev}, P.~S.},
        title = "{Further evidence of quasi-periodic eruptions in a tidal disruption event AT2019vcb by SRG/eROSITA}",
      journal = {\mnras},
         year = 2025,
        month = jun,
       volume = {540},
       number = {1},
        pages = {30-36},
          doi = {10.1093/mnras/staf686},
archivePrefix = {arXiv},
       eprint = {2409.16908},
 primaryClass = {astro-ph.HE},
       adsurl = {https://ui.adsabs.harvard.edu/abs/2025MNRAS.540...30B}
}

@ARTICLE{Allievi2026,
       author = {{Allievi}, Chiara Maria and {Broggi}, Luca and {Sesana}, Alberto and {Bonetti}, Matteo},
        title = "{Two-body relaxation in the EMRI-TDE disk model for Quasi Periodic Eruptions}",
      journal = {arXiv e-prints},
         year = 2026,
        month = mar,
          eid = {arXiv:2603.02302},
        pages = {arXiv:2603.02302},
          doi = {10.48550/arXiv.2603.02302},
archivePrefix = {arXiv},
       eprint = {2603.02302},
 primaryClass = {astro-ph.HE},
       adsurl = {https://ui.adsabs.harvard.edu/abs/2026arXiv260302302A}
}

@ARTICLE{Guo2026,
       author = {{Guo}, Hengxiao and {Yan}, Zhen and {Li}, Ya-Ping and {Chakraborty}, Joheen and {S{\'a}nchez-S{\'a}ez}, Paula and {Hern{\'a}ndez-Garc{\'\i}a}, Lorena and {Zhang}, Wenda and {Sun}, Jingbo and {Li}, Shuang-Liang and {Deng}, Hongping and {Zuo}, Wenwen and {Tagawa}, Hiromichi and {Pan}, Xin and {Zhang}, Minghao and {Ar{\'e}valo}, Patricia and {Lira}, Paulina and {Jin}, Chichuan and {Gu}, Minfeng},
        title = "{Evidence for a Delayed Ultraviolet Counterpart to X-Ray Quasiperiodic Eruptions in Ansky}",
      journal = {\apjl},
         year = 2026,
        month = apr,
       volume = {1000},
       number = {2},
          eid = {L57},
        pages = {L57},
          doi = {10.3847/2041-8213/ae524b},
archivePrefix = {arXiv},
       eprint = {2603.02517},
 primaryClass = {astro-ph.HE},
       adsurl = {https://ui.adsabs.harvard.edu/abs/2026ApJ..1000L..57G}
}

@ARTICLE{Huang2025,
       author = {{Huang}, Xiaoshan and {Linial}, Itai and {Jiang}, Yan-Fei},
        title = "{Multiband Emission from Star─Disk Collisions and Implications for Quasiperiodic Eruptions}",
      journal = {\apj},
         year = 2025,
        month = nov,
       volume = {993},
       number = {2},
          eid = {186},
        pages = {186},
          doi = {10.3847/1538-4357/ae07ca},
archivePrefix = {arXiv},
       eprint = {2506.11231},
 primaryClass = {astro-ph.HE},
       adsurl = {https://ui.adsabs.harvard.edu/abs/2025ApJ...993..186H}
}

@ARTICLE{LiuKun2026,
       author = {{Liu}, Kun and {Liu}, Shang-Fei and {Pan}, Zhen and {Deng}, Hongping and {Shen}, Rongfeng and {Yu}, Cong},
        title = "{Quasi-periodic Eruptions from Stellar-mass Black Holes Impacting Accretion Disks in Galactic Nuclei}",
      journal = {arXiv e-prints},
         year = 2026,
        month = feb,
          eid = {arXiv:2603.00226},
        pages = {arXiv:2603.00226},
          doi = {10.48550/arXiv.2603.00226},
archivePrefix = {arXiv},
       eprint = {2603.00226},
 primaryClass = {astro-ph.HE},
       adsurl = {https://ui.adsabs.harvard.edu/abs/2026arXiv260300226L}
}

@ARTICLE{Huang2026,
       author = {{Huang}, Shunquan and {Huang}, Xiaoshan and {Zhu}, Zhaohuan and {Martin}, Rebecca G.},
        title = "{Resolving Oblique Star-Disk Collisions in Quasi-Periodic Eruptions: Numerical Requirements and the Importance of Geometry}",
      journal = {arXiv e-prints},
         year = 2026,
        month = apr,
          eid = {arXiv:2604.00953},
        pages = {arXiv:2604.00953},
          doi = {10.48550/arXiv.2604.00953},
archivePrefix = {arXiv},
       eprint = {2604.00953},
 primaryClass = {astro-ph.HE},
       adsurl = {https://ui.adsabs.harvard.edu/abs/2026arXiv260400953H}
}

@ARTICLE{Jankovic2026,
       author = {{Jankovi{\v{c}}}, Taj and {Bonnerot}, Cl{\'e}ment and {Karpov}, Sergey and {Jurca}, Aleksej},
        title = "{Radiation-hydrodynamics of star-disc collisions for quasi-periodic eruptions}",
      journal = {arXiv e-prints},
         year = 2026,
        month = feb,
          eid = {arXiv:2602.02656},
        pages = {arXiv:2602.02656},
          doi = {10.48550/arXiv.2602.02656},
archivePrefix = {arXiv},
       eprint = {2602.02656},
 primaryClass = {astro-ph.HE},
       adsurl = {https://ui.adsabs.harvard.edu/abs/2026arXiv260202656J}
}

@ARTICLE{Linial2025,
       author = {{Linial}, Itai and {Metzger}, Brian D. and {Quataert}, Eliot},
        title = "{QPEs from EMRI Debris Streams Impacting Accretion Disks in Galactic Nuclei}",
      journal = {\apj},
         year = 2025,
        month = oct,
       volume = {991},
       number = {2},
          eid = {147},
        pages = {147},
          doi = {10.3847/1538-4357/adfa0e},
archivePrefix = {arXiv},
       eprint = {2506.10096},
 primaryClass = {astro-ph.HE},
       adsurl = {https://ui.adsabs.harvard.edu/abs/2025ApJ...991..147L}
}

@ARTICLE{Chakraborty2025_stime,
       author = {{Chakraborty}, Joheen and {Drummond}, Lisa V. and {Bonetti}, Matteo and {Franchini}, Alessia and {Kejriwal}, Shubham and {Miniutti}, Giovanni and {Arcodia}, Riccardo and {Hughes}, Scott A. and {Duque}, Francisco and {Kara}, Erin and {Sesana}, Alberto and {Giustini}, Margherita and {Motta}, Amedeo and {Burdge}, Kevin},
        title = "{Prospects for EMRI/MBH Parameter Estimation Using Quasiperiodic Eruption Timings: Short-timescale Analysis}",
      journal = {\apj},
         year = 2025,
        month = oct,
       volume = {992},
       number = {1},
          eid = {120},
        pages = {120},
          doi = {10.3847/1538-4357/ae003b},
archivePrefix = {arXiv},
       eprint = {2508.20162},
 primaryClass = {astro-ph.HE},
       adsurl = {https://ui.adsabs.harvard.edu/abs/2025ApJ...992..120C}
}

@ARTICLE{HernandezGarcia2025,
       author = {{Hern{\'a}ndez-Garc{\'\i}a}, L. and {S{\'a}nchez-S{\'a}ez}, P. and {Chakraborty}, J. and {Cuadra}, J. and {Miniutti}, G. and {Arcodia}, R. and {Ar{\'e}valo}, P. and {Giustini}, M. and {Kara}, E. and {Ricci}, C. and {Pasham}, D.~R. and {Arzoumanian}, Z. and {Gendreau}, K. and {Lira}, P.},
        title = "{NICER observations reveal doubled timescales in Ansky's quasi-periodic eruptions}",
      journal = {\aap},
         year = 2025,
        month = nov,
       volume = {703},
          eid = {A263},
        pages = {A263},
          doi = {10.1051/0004-6361/202555258},
archivePrefix = {arXiv},
       eprint = {2509.16304},
 primaryClass = {astro-ph.HE},
       adsurl = {https://ui.adsabs.harvard.edu/abs/2025A&A...703A.263H}
}

@ARTICLE{Chakraborty2026,
       author = {{Chakraborty}, Joheen and {Rappaport}, Saul A. and {Arcodia}, Riccardo and {Linial}, Itai and {Miniutti}, Giovanni and {Burdge}, Kevin B. and {Cuadra}, Jorge and {Giustini}, Margherita and {Hern{\'a}ndez-Garc{\'\i}a}, Lorena and {Kara}, Erin and {Ricci}, Claudio and {S{\'a}nchez-S{\'a}ez}, Paula and {Yao}, Philippe},
        title = "{A Positive Period Derivative in the Quasiperiodic Eruptions of ZTF19acnskyy}",
      journal = {\apjl},
         year = 2026,
        month = apr,
       volume = {1001},
       number = {1},
          eid = {L6},
        pages = {L6},
          doi = {10.3847/2041-8213/ae548b},
archivePrefix = {arXiv},
       eprint = {2602.16776},
 primaryClass = {astro-ph.HE},
       adsurl = {https://ui.adsabs.harvard.edu/abs/2026ApJ..1001L...6C}
}

@ARTICLE{Arcodia2026,
       author = {{Arcodia}, R. and {Miniutti}, G. and {Chakraborty}, J. and {Franchini}, A. and {Giustini}, M. and {Linial}, I. and {Mummery}, A. and {Bertassi}, L. and {Bonetti}, M. and {Kara}, E. and {Merloni}, A. and {Motta}, A. and {Ponti}, G. and {Quintin}, E. and {Soria}, R. and {Baldini}, P. and {Buchner}, J. and {Dotti}, M. and {Fragile}, P.~C. and {Ingram}, A. and {Middleton}, M. and {Panagiotou}, C. and {Sesana}, A. and {Yao}, P. and {Rau}, A. and {Vincentelli}, F.~M. and {Guolo}, M. and {Saxton}, R.},
        title = "{Even a precessing clock is right twice per orbit -- The super-periods of eRO-QPE2 and challenges for quasi-periodic eruption orbital models}",
      journal = {arXiv e-prints},
         year = 2026,
        month = apr,
          eid = {arXiv:2604.09788},
        pages = {arXiv:2604.09788},
          doi = {10.48550/arXiv.2604.09788},
archivePrefix = {arXiv},
       eprint = {2604.09788},
 primaryClass = {astro-ph.HE},
       adsurl = {https://ui.adsabs.harvard.edu/abs/2026arXiv260409788A}
}

@ARTICLE{Zhu2025,
       author = {{Zhu}, Jiazheng and {Jiang}, Ning and {Wang}, Yibo and {Wang}, Tinggui and {Sun}, Luming and {Zhong}, Shiyan and {Yao}, Yuhan and {Chornock}, Ryan and {Dai}, Lixin and {Lyu}, Jianwei and {Shu}, Xinwen and {Fremling}, Christoffer and {Hammerstein}, Erica and {Huang}, Shifeng and {Li}, Wenkai and {You}, Bei},
        title = "{Ultraviolet Spectral Evidence for Ansky as a Slowly Evolving Featureless Tidal Disruption Event with Quasiperiodic Eruptions}",
      journal = {\apjl},
         year = 2025,
        month = nov,
       volume = {994},
       number = {1},
          eid = {L16},
        pages = {L16},
          doi = {10.3847/2041-8213/ae19ea},
archivePrefix = {arXiv},
       eprint = {2510.22211},
 primaryClass = {astro-ph.HE},
       adsurl = {https://ui.adsabs.harvard.edu/abs/2025ApJ...994L..16Z}
}

@ARTICLE{Chakraborty2025,
       author = {{Chakraborty}, Joheen and {Kara}, Erin and {Arcodia}, Riccardo and {Buchner}, Johannes and {Giustini}, Margherita and {Hern{\'a}ndez-Garc{\'\i}a}, Lorena and {Linial}, Itai and {Masterson}, Megan and {Miniutti}, Giovanni and {Mummery}, Andrew and {Panagiotou}, Christos and {Quintin}, Erwan and {S{\'a}nchez-S{\'a}ez}, Paula},
        title = "{Discovery of Quasiperiodic Eruptions in the Tidal Disruption Event and Extreme Coronal Line Emitter AT2022upj: Implications for the QPE/TDE Fraction and a Connection to ECLEs}",
      journal = {\apjl},
         year = 2025,
        month = apr,
       volume = {983},
       number = {2},
          eid = {L39},
        pages = {L39},
          doi = {10.3847/2041-8213/adc2f8},
archivePrefix = {arXiv},
       eprint = {2503.19013},
 primaryClass = {astro-ph.HE},
       adsurl = {https://ui.adsabs.harvard.edu/abs/2025ApJ...983L..39C}
}

@ARTICLE{Miniutti2024,
       author = {{Miniutti}, Giovanni and {Franchini}, Alessia and {Bonetti}, Matteo and {Giustini}, Margherita and {Chakraborty}, Joheen and {Arcodia}, Riccardo and {Saxton}, Richard and {Quintin}, Erwan and {Kosec}, Peter and {Linial}, Itai and {Sesana}, Alberto},
        title = "{Eppur si muove: Evidence of disc precession or a sub-milliparsec SMBH binary in the QPE-emitting galaxy GSN 069}",
      journal = {arXiv e-prints},
         year = 2024,
        month = nov,
          eid = {arXiv:2411.13460},
        pages = {arXiv:2411.13460},
          doi = {10.48550/arXiv.2411.13460},
archivePrefix = {arXiv},
       eprint = {2411.13460},
 primaryClass = {astro-ph.HE},
       adsurl = {https://ui.adsabs.harvard.edu/abs/2024arXiv241113460M}
}

@ARTICLE{Xie2012,
       author = {{Xie}, Fu-Guo and {Yuan}, Feng},
        title = "{Radiative efficiency of hot accretion flows}",
      journal = {\mnras},
         year = 2012,
        month = dec,
       volume = {427},
       number = {2},
        pages = {1580-1586},
          doi = {10.1111/j.1365-2966.2012.22030.x},
archivePrefix = {arXiv},
       eprint = {1207.3113},
 primaryClass = {astro-ph.HE},
       adsurl = {https://ui.adsabs.harvard.edu/abs/2012MNRAS.427.1580X}
}

@ARTICLE{Tagawa2023_SC,
       author = {{Tagawa}, Hiromichi and {Kimura}, Shigeo S. and {Haiman}, Zolt{\'a}n and {Perna}, Rosalba and {Bartos}, Imre},
        title = "{Shock cooling and breakout emission for optical flares associated with gravitational wave events}",
      journal = {arXiv e-prints},
         year = 2023,
        month = oct,
          eid = {arXiv:2310.18392},
        pages = {arXiv:2310.18392},
          doi = {10.48550/arXiv.2310.18392},
archivePrefix = {arXiv},
       eprint = {2310.18392},
 primaryClass = {astro-ph.HE},
       adsurl = {https://ui.adsabs.harvard.edu/abs/2023arXiv231018392T}
}

@ARTICLE{Launhardt2002,
       author = {{Launhardt}, R. and {Zylka}, R. and {Mezger}, P.~G.},
        title = "{The nuclear bulge of the Galaxy. III. Large-scale physical characteristics of stars and interstellar matter}",
      journal = {\aap},
         year = 2002,
        month = mar,
       volume = {384},
        pages = {112-139},
          doi = {10.1051/0004-6361:20020017},
archivePrefix = {arXiv},
       eprint = {astro-ph/0201294},
 primaryClass = {astro-ph},
       adsurl = {https://ui.adsabs.harvard.edu/abs/2002A&A...384..112L}
}

@ARTICLE{Chevalier2006,
       author = {{Chevalier}, Roger A. and {Fransson}, Claes},
        title = "{Circumstellar Emission from Type Ib and Ic Supernovae}",
      journal = {\apj},
         year = 2006,
        month = nov,
       volume = {651},
       number = {1},
        pages = {381-391},
          doi = {10.1086/507606},
archivePrefix = {arXiv},
       eprint = {astro-ph/0607196},
 primaryClass = {astro-ph},
       adsurl = {https://ui.adsabs.harvard.edu/abs/2006ApJ...651..381C}
}

@ARTICLE{Soderberg2012,
       author = {{Soderberg}, A.~M. and {Margutti}, R. and {Zauderer}, B.~A. and {Krauss}, M. and {Katz}, B. and {Chomiuk}, L. and {Dittmann}, J.~A. and {Nakar}, E. and {Sakamoto}, T. and {Kawai}, N. and {Hurley}, K. and {Barthelmy}, S. and {Toizumi}, T. and {Morii}, M. and {Chevalier}, R.~A. and {Gurwell}, M. and {Petitpas}, G. and {Rupen}, M. and {Alexander}, K.~D. and {Levesque}, E.~M. and {Fransson}, C. and {Brunthaler}, A. and {Bietenholz}, M.~F. and {Chugai}, N. and {Grindlay}, J. and {Copete}, A. and {Connaughton}, V. and {Briggs}, M. and {Meegan}, C. and {von Kienlin}, A. and {Zhang}, X. and {Rau}, A. and {Golenetskii}, S. and {Mazets}, E. and {Cline}, T.},
        title = "{Panchromatic Observations of SN 2011dh Point to a Compact Progenitor Star}",
      journal = {\apj},
         year = 2012,
        month = jun,
       volume = {752},
       number = {2},
          eid = {78},
        pages = {78},
          doi = {10.1088/0004-637X/752/2/78},
archivePrefix = {arXiv},
       eprint = {1107.1876},
 primaryClass = {astro-ph.HE},
       adsurl = {https://ui.adsabs.harvard.edu/abs/2012ApJ...752...78S}
}

@ARTICLE{Chevalier1982,
       author = {{Chevalier}, R.~A.},
        title = "{Self-similar solutions for the interaction of stellar ejecta with an external medium.}",
      journal = {\apj},
         year = 1982,
        month = jul,
       volume = {258},
        pages = {790-797},
          doi = {10.1086/160126},
       adsurl = {https://ui.adsabs.harvard.edu/abs/1982ApJ...258..790C}
}

@ARTICLE{Matsuoka2024,
       author = {{Matsuoka}, Tomoki and {Kimura}, Shigeo S. and {Maeda}, Keiichi and {Tanaka}, Masaomi},
        title = "{A New Insight into Electron Acceleration Properties from Theoretical Modeling of Double-peaked Radio Light Curves in Core-collapse Supernovae}",
      journal = {\apj},
         year = 2024,
        month = jan,
       volume = {960},
       number = {1},
          eid = {70},
        pages = {70},
          doi = {10.3847/1538-4357/ad096c},
archivePrefix = {arXiv},
       eprint = {2309.01209},
 primaryClass = {astro-ph.HE},
       adsurl = {https://ui.adsabs.harvard.edu/abs/2024ApJ...960...70M}
}

@ARTICLE{Pasham2024_Swift,
       author = {{Pasham}, Dheeraj and {Coughlin}, Eric and {Nixon}, Chris and {Zajacek}, Michal and {Sukova}, Petra and {Karas}, Vladimir and {Wevers}, Thomas and {Tombesi}, Francesco},
        title = "{Repeated Partial Tidal Disruptions and Quasi-Periodic Eruptions in SwJ023017.0+283603}",
      journal = {arXiv e-prints},
         year = 2024,
        month = nov,
          eid = {arXiv:2411.05948},
        pages = {arXiv:2411.05948},
archivePrefix = {arXiv},
       eprint = {2411.05948},
 primaryClass = {astro-ph.HE},
       adsurl = {https://ui.adsabs.harvard.edu/abs/2024arXiv241105948P}
}

@ARTICLE{Fragner2010,
       author = {{Fragner}, M.~M. and {Nelson}, R.~P.},
        title = "{Evolution of warped and twisted accretion discs in close binary systems}",
      journal = {\aap},
         year = 2010,
        month = feb,
       volume = {511},
          eid = {A77},
        pages = {A77},
          doi = {10.1051/0004-6361/200913088},
archivePrefix = {arXiv},
       eprint = {0912.3220},
 primaryClass = {astro-ph.EP},
       adsurl = {https://ui.adsabs.harvard.edu/abs/2010A&A...511A..77F}
}

@ARTICLE{Wijers1999,
       author = {{Wijers}, Ralph A.~M.~J. and {Pringle}, J.~E.},
        title = "{Warped accretion discs and the long periods in X-ray binaries}",
      journal = {\mnras},
         year = 1999,
        month = sep,
       volume = {308},
       number = {1},
        pages = {207-220},
          doi = {10.1046/j.1365-8711.1999.02720.x},
archivePrefix = {arXiv},
       eprint = {astro-ph/9811056},
 primaryClass = {astro-ph},
       adsurl = {https://ui.adsabs.harvard.edu/abs/1999MNRAS.308..207W}
}

@ARTICLE{WangY2024_capture,
       author = {{Wang}, Yihan and {Zhu}, Zhaohuan and {Lin}, Douglas N.~C.},
        title = "{Stellar/BH population in AGN discs: direct binary formation from capture objects in nuclei clusters}",
      journal = {\mnras},
         year = 2024,
        month = mar,
       volume = {528},
       number = {3},
        pages = {4958-4975},
          doi = {10.1093/mnras/stae321},
archivePrefix = {arXiv},
       eprint = {2308.09129},
 primaryClass = {astro-ph.GA},
       adsurl = {https://ui.adsabs.harvard.edu/abs/2024MNRAS.528.4958W}
}

@ARTICLE{Arcodia2024_QPE2,
       author = {{Arcodia}, R. and {Linial}, I. and {Miniutti}, G. and {Franchini}, A. and {Giustini}, M. and {Bonetti}, M. and {Sesana}, A. and {Soria}, R. and {Chakraborty}, J. and {Dotti}, M. and {Kara}, E. and {Merloni}, A. and {Ponti}, G. and {Vincentelli}, F.},
        title = "{Ticking away: The long-term X-ray timing and spectral evolution of eRO-QPE2}",
      journal = {\aap},
         year = 2024,
        month = oct,
       volume = {690},
          eid = {A80},
        pages = {A80},
          doi = {10.1051/0004-6361/202451218},
archivePrefix = {arXiv},
       eprint = {2406.17020},
 primaryClass = {astro-ph.HE},
       adsurl = {https://ui.adsabs.harvard.edu/abs/2024A&A...690A..80A}
}

@ARTICLE{WangY2024_TDE_AGN,
       author = {{Wang}, Yihan and {Lin}, Douglas N.~C. and {Zhang}, Bing and {Zhu}, Zhaohuan},
        title = "{Changing-look Active Galactic Nuclei Behavior Induced by Disk-captured Tidal Disruption Events}",
      journal = {\apjl},
         year = 2024,
        month = feb,
       volume = {962},
       number = {1},
          eid = {L7},
        pages = {L7},
          doi = {10.3847/2041-8213/ad20e5},
archivePrefix = {arXiv},
       eprint = {2310.00038},
 primaryClass = {astro-ph.HE},
       adsurl = {https://ui.adsabs.harvard.edu/abs/2024ApJ...962L...7W}
}

@ARTICLE{Chen2024_pop,
       author = {{Chen}, Yi-Xian and {Lin}, Douglas N.~C.},
        title = "{The Population of Massive Stars in Active Galactic Nuclei Disks}",
      journal = {\apj},
         year = 2024,
        month = jun,
       volume = {967},
       number = {2},
          eid = {88},
        pages = {88},
          doi = {10.3847/1538-4357/ad3c3a},
archivePrefix = {arXiv},
       eprint = {2404.08780},
 primaryClass = {astro-ph.GA},
       adsurl = {https://ui.adsabs.harvard.edu/abs/2024ApJ...967...88C}
}

@ARTICLE{Zhou2024b,
       author = {{Zhou}, Cong and {Zhong}, Binyu and {Zeng}, Yuhe and {Huang}, Lei and {Pan}, Zhen},
        title = "{Probing orbits of stellar mass objects deep in galactic nuclei with quasiperiodic eruptions. II. Population analysis}",
      journal = {\prd},
         year = 2024,
        month = oct,
       volume = {110},
       number = {8},
          eid = {083019},
        pages = {083019},
          doi = {10.1103/PhysRevD.110.083019},
archivePrefix = {arXiv},
       eprint = {2405.06429},
 primaryClass = {astro-ph.HE},
       adsurl = {https://ui.adsabs.harvard.edu/abs/2024PhRvD.110h3019Z}
}

@ARTICLE{Gilbert2024,
       author = {{Gilbert}, Olivier and {Ruan}, John J. and {Eracleous}, Michael and {Haggard}, Daryl and {Runnoe}, Jessie C.},
        title = "{A Host Galaxy Morphology Link Between Quasi-Periodic Eruptions and Tidal Disruption Events}",
      journal = {arXiv e-prints},
         year = 2024,
        month = sep,
          eid = {arXiv:2409.10486},
        pages = {arXiv:2409.10486},
          doi = {10.48550/arXiv.2409.10486},
archivePrefix = {arXiv},
       eprint = {2409.10486},
 primaryClass = {astro-ph.HE},
       adsurl = {https://ui.adsabs.harvard.edu/abs/2024arXiv240910486G}
}

@ARTICLE{Vurm2024,
       author = {{Vurm}, Indrek and {Linial}, Itai and {Metzger}, Brian D.},
        title = "{Radiation Transport Simulations of Quasi-Periodic Eruptions from Star-Disk Collisions}",
      journal = {arXiv e-prints},
         year = 2024,
        month = oct,
          eid = {arXiv:2410.05166},
        pages = {arXiv:2410.05166},
          doi = {10.48550/arXiv.2410.05166},
archivePrefix = {arXiv},
       eprint = {2410.05166},
 primaryClass = {astro-ph.HE},
       adsurl = {https://ui.adsabs.harvard.edu/abs/2024arXiv241005166V}
}

@ARTICLE{Kejriwal2024,
       author = {{Kejriwal}, Shubham and {Witzany}, Vojt{\v{e}}ch and {Zaja{\v{c}}ek}, Michal and {Pasham}, Dheeraj R. and {Chua}, Alvin J.~K.},
        title = "{Repeating nuclear transients as candidate electromagnetic counterparts of LISA extreme mass ratio inspirals}",
      journal = {\mnras},
         year = 2024,
        month = aug,
       volume = {532},
       number = {2},
        pages = {2143-2158},
          doi = {10.1093/mnras/stae1599},
archivePrefix = {arXiv},
       eprint = {2404.00941},
 primaryClass = {astro-ph.HE},
       adsurl = {https://ui.adsabs.harvard.edu/abs/2024MNRAS.532.2143K}
}

@ARTICLE{Wevers2024,
       author = {{Wevers}, T. and {French}, K.~D. and {Zabludoff}, A.~I. and {Fischer}, T.~C. and {Rowlands}, K. and {Guolo}, M. and {Dalla Barba}, B. and {Arcodia}, R. and {Berton}, M. and {Bian}, F. and {Linial}, I. and {Miniutti}, G. and {Pasham}, D.~R.},
        title = "{X-Ray Quasi-periodic Eruptions and Tidal Disruption Events Prefer Similar Host Galaxies}",
      journal = {\apjl},
         year = 2024,
        month = jul,
       volume = {970},
       number = {1},
          eid = {L23},
        pages = {L23},
          doi = {10.3847/2041-8213/ad5f1b},
archivePrefix = {arXiv},
       eprint = {2406.02678},
 primaryClass = {astro-ph.HE},
       adsurl = {https://ui.adsabs.harvard.edu/abs/2024ApJ...970L..23W}
}

@ARTICLE{Yao2024,
       author = {{Yao}, Philippe Z. and {Quataert}, Eliot and {Jiang}, Yan-Fei and {Lu}, Wenbin and {White}, Christopher J.},
        title = "{Star-Disk Collisions: Implications for QPEs and Other Transients Near Supermassive Black Holes}",
      journal = {arXiv e-prints},
         year = 2024,
        month = jul,
          eid = {arXiv:2407.14578},
        pages = {arXiv:2407.14578},
          doi = {10.48550/arXiv.2407.14578},
archivePrefix = {arXiv},
       eprint = {2407.14578},
 primaryClass = {astro-ph.HE},
       adsurl = {https://ui.adsabs.harvard.edu/abs/2024arXiv240714578Y}
}

@ARTICLE{Kaur2024_FP,
       author = {{Kaur}, Karamveer and {Rom}, Barak and {Sari}, Re'em},
        title = "{Semi-Analytical Fokker Planck Models for Nuclear Star Clusters}",
      journal = {arXiv e-prints},
         year = 2024,
        month = jun,
          eid = {arXiv:2406.07627},
        pages = {arXiv:2406.07627},
          doi = {10.48550/arXiv.2406.07627},
archivePrefix = {arXiv},
       eprint = {2406.07627},
 primaryClass = {astro-ph.HE},
       adsurl = {https://ui.adsabs.harvard.edu/abs/2024arXiv240607627K}
}

@ARTICLE{Chakraborty2024,
       author = {{Chakraborty}, Joheen and {Arcodia}, Riccardo and {Kara}, Erin and {Miniutti}, Giovanni and {Giustini}, Margherita and {Tetarenko}, Alexandra J. and {Rhodes}, Lauren and {Franchini}, Alessia and {Bonetti}, Matteo and {Burdge}, Kevin B. and {Goodwin}, Adelle J. and {Maccarone}, Thomas J. and {Merloni}, Andrea and {Ponti}, Gabriele and {Remillard}, Ronald A. and {Saxton}, Richard D.},
        title = "{Testing EMRI Models for Quasi-periodic Eruptions with 3.5 yr of Monitoring eRO-QPE1}",
      journal = {\apj},
         year = 2024,
        month = apr,
       volume = {965},
       number = {1},
          eid = {12},
        pages = {12},
          doi = {10.3847/1538-4357/ad2941},
archivePrefix = {arXiv},
       eprint = {2402.08722},
 primaryClass = {astro-ph.HE},
       adsurl = {https://ui.adsabs.harvard.edu/abs/2024ApJ...965...12C}
}

@ARTICLE{Zhou2024a,
       author = {{Zhou}, Cong and {Huang}, Lei and {Guo}, Kangrou and {Li}, Ya-Ping and {Pan}, Zhen},
        title = "{Probing orbits of stellar mass objects deep in galactic nuclei with quasiperiodic eruptions}",
      journal = {\prd},
         year = 2024,
        month = may,
       volume = {109},
       number = {10},
          eid = {103031},
        pages = {103031},
          doi = {10.1103/PhysRevD.109.103031},
archivePrefix = {arXiv},
       eprint = {2401.11190},
 primaryClass = {astro-ph.HE},
       adsurl = {https://ui.adsabs.harvard.edu/abs/2024PhRvD.109j3031Z}
}

@ARTICLE{Ressler2024,
       author = {{Ressler}, Sean M. and {Combi}, Luciano and {Li}, Xinyu and {Ripperda}, Bart and {Yang}, Huan},
        title = "{Black Hole{\textendash}Disk Interactions in Magnetically Arrested Active Galactic Nuclei: General Relativistic Magnetohydrodynamic Simulations Using a Time-dependent, Binary Metric}",
      journal = {\apj},
         year = 2024,
        month = may,
       volume = {967},
       number = {1},
          eid = {70},
        pages = {70},
          doi = {10.3847/1538-4357/ad3ae2},
archivePrefix = {arXiv},
       eprint = {2404.02193},
 primaryClass = {astro-ph.HE},
       adsurl = {https://ui.adsabs.harvard.edu/abs/2024ApJ...967...70R}
}

@ARTICLE{Arcodia2024,
       author = {{Arcodia}, R. and {Merloni}, A. and {Buchner}, J. and {Baldini}, P. and {Ponti}, G. and {Rau}, A. and {Liu}, Z. and {Nandra}, K. and {Salvato}, M.},
        title = "{Cosmic hide and seek: The volumetric rate of X-ray quasi-periodic eruptions}",
      journal = {\aap},
         year = 2024,
        month = apr,
       volume = {684},
          eid = {L14},
        pages = {L14},
          doi = {10.1051/0004-6361/202348949},
archivePrefix = {arXiv},
       eprint = {2403.17059},
 primaryClass = {astro-ph.HE},
       adsurl = {https://ui.adsabs.harvard.edu/abs/2024A&A...684L..14A}
}

@ARTICLE{Miniutti2023_reb,
       author = {{Miniutti}, G. and {Giustini}, M. and {Arcodia}, R. and {Saxton}, R.~D. and {Chakraborty}, J. and {Read}, A.~M. and {Kara}, E.},
        title = "{Alive and kicking: A new QPE phase in GSN 069 revealing a quiescent luminosity threshold for QPEs}",
      journal = {\aap},
         year = 2023,
        month = jun,
       volume = {674},
          eid = {L1},
        pages = {L1},
          doi = {10.1051/0004-6361/202346653},
archivePrefix = {arXiv},
       eprint = {2305.09717},
 primaryClass = {astro-ph.HE},
       adsurl = {https://ui.adsabs.harvard.edu/abs/2023A&A...674L...1M}
}

@ARTICLE{Tagawa2023,
       author = {{Tagawa}, Hiromichi and {Haiman}, Zolt{\'a}n},
        title = "{Flares from stars crossing active galactic nucleus discs on low-inclination orbits}",
      journal = {\mnras},
         year = 2023,
        month = nov,
       volume = {526},
       number = {1},
        pages = {69-79},
          doi = {10.1093/mnras/stad2616},
archivePrefix = {arXiv},
       eprint = {2304.03670},
 primaryClass = {astro-ph.HE},
       adsurl = {https://ui.adsabs.harvard.edu/abs/2023MNRAS.526...69T}
}

@ARTICLE{Nicholl2024,
       author = {{Nicholl}, M. and {Pasham}, D.~R. and {Mummery}, A. and {Guolo}, M. and {Gendreau}, K. and {Dewangan}, G.~C. and {Ferrara}, E.~C. and {Remillard}, R. and {Bonnerot}, C. and {Chakraborty}, J. and {Hajela}, A. and {Dhillon}, V.~S. and {Gillan}, A.~F. and {Greenwood}, J. and {Huber}, M.~E. and {Janiuk}, A. and {Salvesen}, G. and {van Velzen}, S. and {Aamer}, A. and {Alexander}, K.~D. and {Angus}, C.~R. and {Arzoumanian}, Z. and {Auchettl}, K. and {Berger}, E. and {de Boer}, T. and {Cendes}, Y. and {Chambers}, K.~C. and {Chen}, T. -W. and {Chornock}, R. and {Fulton}, M.~D. and {Gao}, H. and {Gillanders}, J.~H. and {Gomez}, S. and {Gompertz}, B.~P. and {Fabian}, A.~C. and {Herman}, J. and {Ingram}, A. and {Kara}, E. and {Laskar}, T. and {Lawrence}, A. and {Lin}, C. -C. and {Lowe}, T.~B. and {Magnier}, E.~A. and {Margutti}, R. and {McGee}, S.~L. and {Minguez}, P. and {Moore}, T. and {Nathan}, E. and {Oates}, S.~R. and {Patra}, K.~C. and {Ramsden}, P. and {Ravi}, V. and {Ridley}, E.~J. and {Sheng}, X. and {Smartt}, S.~J. and {Smith}, K.~W. and {Srivastav}, S. and {Stein}, R. and {Stevance}, H.~F. and {Turner}, S.~G.~D. and {Wainscoat}, R.~J. and {Weston}, J. and {Wevers}, T. and {Young}, D.~R.},
        title = "{Quasi-periodic X-ray eruptions years after a nearby tidal disruption event}",
      journal = {\nat},
         year = 2024,
        month = oct,
       volume = {634},
       number = {8035},
        pages = {804-808},
          doi = {10.1038/s41586-024-08023-6},
archivePrefix = {arXiv},
       eprint = {2409.02181},
 primaryClass = {astro-ph.HE},
       adsurl = {https://ui.adsabs.harvard.edu/abs/2024Natur.634..804N}
}

@ARTICLE{Guolo2024,
       author = {{Guolo}, Muryel and {Pasham}, Dheeraj R. and {Zaja{\v{c}}ek}, Michal and {Coughlin}, Eric R. and {Gezari}, Suvi and {Sukov{\'a}}, Petra and {Wevers}, Thomas and {Witzany}, Vojt{\v{e}}ch and {Tombesi}, Francesco and {van Velzen}, Sjoert and {Alexander}, Kate D. and {Yao}, Yuhan and {Arcodia}, Riccardo and {Karas}, Vladim{\'\i}r and {Miller-Jones}, James C.~A. and {Remillard}, Ronald and {Gendreau}, Keith and {Ferrara}, Elizabeth C.},
        title = "{X-ray eruptions every 22 days from the nucleus of a nearby galaxy}",
      journal = {Nature Astronomy},
         year = 2024,
        month = mar,
       volume = {8},
        pages = {347-358},
          doi = {10.1038/s41550-023-02178-4},
archivePrefix = {arXiv},
       eprint = {2309.03011},
 primaryClass = {astro-ph.HE},
       adsurl = {https://ui.adsabs.harvard.edu/abs/2024NatAs...8..347G}
}

@ARTICLE{Arcodia2024_QPE34,
       author = {{Arcodia}, R. and {Liu}, Z. and {Merloni}, A. and {Malyali}, A. and {Rau}, A. and {Chakraborty}, J. and {Goodwin}, A. and {Buckley}, D. and {Brink}, J. and {Gromadzki}, M. and {Arzoumanian}, Z. and {Buchner}, J. and {Kara}, E. and {Nandra}, K. and {Ponti}, G. and {Salvato}, M. and {Anderson}, G. and {Baldini}, P. and {Grotova}, I. and {Krumpe}, M. and {Maitra}, C. and {Miller-Jones}, J.~C.~A. and {Ramos-Ceja}, M.~E.},
        title = "{The more the merrier: SRG/eROSITA discovers two further galaxies showing X-ray quasi-periodic eruptions}",
      journal = {\aap},
         year = 2024,
        month = apr,
       volume = {684},
          eid = {A64},
        pages = {A64},
          doi = {10.1051/0004-6361/202348881},
archivePrefix = {arXiv},
       eprint = {2401.17275},
 primaryClass = {astro-ph.HE},
       adsurl = {https://ui.adsabs.harvard.edu/abs/2024A&A...684A..64A}
}

@ARTICLE{Natarajan1998,
       author = {{Natarajan}, Priyamvada and {Pringle}, J.~E.},
        title = "{The Alignment of Disk and Black Hole Spins in Active Galactic Nuclei}",
      journal = {\apjl},
         year = 1998,
        month = oct,
       volume = {506},
       number = {2},
        pages = {L97-L100},
          doi = {10.1086/311658},
archivePrefix = {arXiv},
       eprint = {astro-ph/9808187},
 primaryClass = {astro-ph},
       adsurl = {https://ui.adsabs.harvard.edu/abs/1998ApJ...506L..97N}
}

@ARTICLE{Miniutti2023_reQPE,
       author = {{Miniutti}, G. and {Giustini}, M. and {Arcodia}, R. and {Saxton}, R.~D. and {Chakraborty}, J. and {Read}, A.~M. and {Kara}, E.},
        title = "{Alive and kicking: A new QPE phase in GSN 069 revealing a quiescent luminosity threshold for QPEs}",
      journal = {\aap},
         year = 2023,
        month = jun,
       volume = {674},
          eid = {L1},
        pages = {L1},
          doi = {10.1051/0004-6361/202346653},
archivePrefix = {arXiv},
       eprint = {2305.09717},
 primaryClass = {astro-ph.HE},
       adsurl = {https://ui.adsabs.harvard.edu/abs/2023A&A...674L...1M}
}

@ARTICLE{Quintin2023,
       author = {{Quintin}, E. and {Webb}, N.~A. and {Guillot}, S. and {Miniutti}, G. and {Kammoun}, E.~S. and {Giustini}, M. and {Arcodia}, R. and {Soucail}, G. and {Clerc}, N. and {Amato}, R. and {Markwardt}, C.~B.},
        title = "{Tormund's return: Hints of quasi-periodic eruption features from a recent optical tidal disruption event}",
      journal = {arXiv e-prints},
         year = 2023,
        month = jun,
          eid = {arXiv:2306.00438},
        pages = {arXiv:2306.00438},
          doi = {10.48550/arXiv.2306.00438},
archivePrefix = {arXiv},
       eprint = {2306.00438},
 primaryClass = {astro-ph.HE},
       adsurl = {https://ui.adsabs.harvard.edu/abs/2023arXiv230600438Q}
}

@BOOK{Rybicki1979,
       author = {{Rybicki}, George B. and {Lightman}, Alan P.},
        title = "{Radiative processes in astrophysics}",
         year = 1979,
       adsurl = {https://ui.adsabs.harvard.edu/abs/1979rpa..book.....R},
       publisher = {Wiley and Sons, Incorporated, John}
}

\end{document}